\PassOptionsToPackage{names,dvipsnames}{xcolor} 

\documentclass[conference]{IEEEtran}

\usepackage{mwe}
\usepackage{amsmath}\allowdisplaybreaks
\usepackage{amssymb}
\usepackage{abraces}

\newif\ifedit
\edittrue
\ifedit
\usepackage[colorinlistoftodos]{todonotes}
\else
\usepackage[colorinlistoftodos,disable]{todonotes}
\fi
\usepackage[nounderscore]{syntax}
\usepackage{semantic}
\usepackage{amsfonts,amsthm,xspace}
\usepackage{mathtools}
\usepackage{listings}
\usepackage{pgf}
\usepackage{pgfplots}
\usepackage{filecontents}
\usepackage{tikz}
\usetikzlibrary{positioning,shadows,arrows,calc,backgrounds,fit,shapes,shapes.multipart,decorations.pathreplacing,shapes.misc,patterns,decorations.markings,decorations.pathmorphing}
\tikzset{snake it/.style={decorate, decoration=snake}}
\tikzset{srcredstar/.style={dotted, draw = \stlccol,->}}
\tikzset{trgredstar/.style={dotted, draw = \ulccol,->,thick}}
\tikzset{srcred/.style={, draw = \stlccol, bend left, ->}}
\tikzset{trgred/.style={, draw = \ulccol, bend left, ->}}
\tikzset{staterel/.style={}}
\tikzset{actrel/.style={-,thin, dotted}}
\tikzset{staterelstrong/.style={-,thin, double equal sign distance}}
\tikzset{staterelweak/.style={-,thin, }}
\tikzset{srcstate/.style={draw=\stlccol,rounded corners,font=\footnotesize}}
\tikzset{trgstate/.style={draw=\ulccol,rounded corners,font=\footnotesize}}
\tikzset{myimpl/.style={,double,-implies,double equal sign distance}}
\tikzset{myptr/.style={decoration={markings,mark=at position 1 with {\arrow[scale=2,>=stealth]{>}}},postaction={decorate}}}
\tikzset{canchange/.style={loosely dotted,draw=\ulccol,very thick}}
\tikzset{cannotchange/.style={dashed,draw=\ulccol,thick}}
\tikzset{candiffer/.style={draw=black}}
\tikzset{cannotdiffer/.style={draw=black, double equal sign distance}}
\tikzset{hide/.style={opacity=0}}
\tikzset{semihide/.style={opacity=.5}}
\usepackage{subcaption}
\usepackage{verbatim}
\usepackage[utf8]{inputenc}
\usepackage{parcolumns}
\usepackage{xcolor}
\usepackage[T1]{fontenc}
\usepackage[scaled=.83]{beramono}
\usepackage{hyperref}
\usepackage{cleveref}
\usepackage[most]{tcolorbox}
\usepackage[inline]{enumitem}
\usepackage{multicol}
\usepackage{stmaryrd}
\usepackage[numbers]{natbib}
\usepackage{bm}

\usepackage{wrapfig}
\usepackage{soul}

\hypersetup{ 
	colorlinks,
	linkcolor={ForestGreen!50!black},
	citecolor={blue!50!black},
	urlcolor={CarnationPink!80!black} }

\usetikzlibrary{arrows,matrix}
\newcommand{\pac}[0]{\mi{PAC}\xspace}
\newcommand{\fa}[0]{\mi{FA}\xspace}

\newcommand{\tricl}[0]{\mi{TrICL}\xspace}
\newcommand{\cheri}[0]{\mi{CHERI}\xspace}

\newcommand{\MPin}[1]{\todo[color=blue!30,inline]{MP TODO: #1}}

\newcommand{\AKin}[1]{\todo[color=yellow,inline]{AK: #1}}

\crefname{lstlisting}{Listing}{Listings}
\crefname{typerule}{rule}{rules}
\newcounter{typerule}
\newcommand{\typeruleInt}[5]{
	\def\thetyperule{#1}%
	\refstepcounter{typerule}%
	\label{tr:#4}%
	\ensuremath{\begin{array}{c}#5 \inference{#2}{#3}\end{array}}
}

\newcommand{\typerule}[4]{
	\typeruleInt{#1}{#2}{#3}{#4}{\textsf{\scriptsize ({#1})} \\      }
}

\newcounter{typeruled}

\newcommand{\neutcol}[0]{black}
\newcommand{\stlccol}[0]{RoyalBlue}
\newcommand{\ulccol}[0]{RedOrange}
\newcommand{\commoncol}[0]{black}    

\newcommand{\bl}[1]{\col{\neutcol }{#1}}
\newcommand{\col}[2]{\ensuremath{{\color{#1}{#2}}}}
\newcommand{\src}[1]{\ensuremath{\mathsf{\col{\stlccol}{{#1}}}}}
\newcommand{\trg}[1]{\ensuremath{\mathbf{\col{\ulccol }{{#1}}}}}
\newcommand{\trgb}[1]{\ensuremath{\bm{\col{\ulccol }{#1}}}}
\newcommand{\com}[1]{\mi{\col{\commoncol }{#1}}}
\newcommand{\neutralcol}[1]{\mathit{{\col{\neutcol}{{#1}}}}}

\newcommand{\OB}[1]{\overline{#1}}

\newcommand{\ms}[1]{\mathsf{#1}}

\newcommand{\funname}[1]{\mathtt{#1}}
\newcommand{\fun}[2]{\ensuremath{{\neutralcol{\funname{#1}\left(#2\right)}}}\xspace}

\newcommand{\targetlanguage}[0]{\trg{CHERIExp}\xspace}
\newcommand{\sourcelanguage}[0]{\src{ImpMod}\xspace}

\newcommand{\compilersymbol}[1]{\ensuremath{\bl{\left\llbracket {#1} \right\rrbracket}}}
\newcommand{\instructiontrans}[1]{\ensuremath{\llparenthesis {#1} \rrparenthesis}}
\newcommand{\expressiontrans}[1]{\ensuremath{\lbag {#1} \rbag}}

\newcommand{\defeq}[0]{\ensuremath{\overset{\mathsf{def}}{=}}}

\newcommand{\paronto}[1]{%
	\rightharpoonup\mathrel{\mspace{-15mu}}\xrightharpoonup{#1}
}

\newcommand{\negativeInfinity}{\nabla}

\newcommand{\statecrossrelated}[0]{\ensuremath{\cong}}
\newcommand{\strongsim}[0]{\ensuremath{\approx}}
\newcommand{\weaksim}[0]{\ensuremath{\sim}}

\newtheorem{lemma}{Lemma}
\newtheorem{definition}{Definition}

\newtheorem{fact}{Fact}

\newcommand{\keywordCol}[0]{\textCol}
\newcommand{\textCol}[0]{\stlccol}
\newcommand{\textColtwo}[0]{\ulccol}
\newcommand{\stringCol}[0]{\textCol}
\definecolor{mygray}{rgb}{0.5,0.5,0.5}

\lstdefinelanguage{Java} 
{morekeywords={abstract, all, and, as, assert, but, check, disj, else, exactly, extends, fact, for, fun, iden, if, iff, implies, in, Int, int, let, lone, module, no, none, not, one, open, or, part, pred, run, seq, set, sig, some, sum, then, univ, package, class, public, private, null, return, new, interface, extern, object, implements, System, static, super, try , catch, throw, throws, Unit, var, val, of, principal, trust},
	sensitive=true,
	commentstyle=\color{purple!40!black},
	morecomment=[l][\footnotesize\itshape\color{purple!40!black}]{//},
	keywordstyle=\color{\keywordCol}, 
	identifierstyle=\color{\textCol},
	stringstyle=\color{\stringCol},
	basicstyle={\footnotesize\ttfamily\color{\stringCol}},
	numbers=left,
	numberstyle=\tiny\color{mygray},
	tabsize=2,
	numbersep=3pt,
	breaklines=true,
	lineskip=-2pt,
	stepnumber=1,
	captionpos=b,
	breaklines=true,
	breakatwhitespace=false,
	showspaces=false,
	showtabs=false,
	float=!h,
	columns=fullflexible,escapeinside={(*@}{@*)},
	moredelim=**[is][\color{red!60}]{@}{@},
	literate={->}{{$\to$}}1 {^}{{$\mspace{-3mu}\widehat{\quad}\mspace{-3mu}$}}1
	{<}{$<$ }2 {>}{$>$ }2 {>=}{$\geq$ }2 {=<}{$\leq$ }2
	{<:}{{$<\mspace{-3mu}:$}}2 {:>}{{$:\mspace{-3mu}>$}}2
	{<=}{{$\Leftarrow$ }}2
	{=>}{{$\Rightarrow$ }}2 {+}{$+$ }2 {++}{{$+\mspace{-8mu}+$ }}2
	{<=>}{{$\Leftrightarrow$ }}2 {+}{$+$ }2 {++}{{$+\mspace{-8mu}+$ }}2
	{\~}{{$\mspace{-3mu}\widetilde{\quad}\mspace{-3mu}$}}1
	{!=}{$\neq$ }2 {*}{${}^{\ast}$}1 
	{\#}{$\#$}1
}
\lstdefinelanguage{Ass} 
{morekeywords={abstract, all, and, as, assert, but, check, disj, else, exactly, extends, fact, for, fun, iden, if, iff, implies, in, Int, int, let, lone, module, no, none, not, one, open, or, part, pred, run, seq, set, sig, some, sum, then, univ, package, class, public, private, null, return, new, interface, extern, object, implements, System, static, super, try , catch, throw, throws, Unit, var, val, of, principal, trust},
	sensitive=true,
	commentstyle=\color{purple!40!black},
	morecomment=[l][\footnotesize\itshape\color{purple!40!black}]{//},
	identifierstyle=\color{\textColtwo},
	basicstyle={\footnotesize\ttfamily\bfseries\color{\textColtwo}},
	numbers=left,
	numberstyle=\tiny\color{mygray},
	tabsize=2,
	numbersep=3pt,
	breaklines=true,
	lineskip=-2pt,
	stepnumber=1,
	captionpos=b,
	breaklines=true,
	breakatwhitespace=false,
	showspaces=false,
	showtabs=false,
	float=!h,
	columns=fullflexible,escapeinside={(*@}{@*)},
	moredelim=**[is][\color{red!60}]{@}{@},
	literate={->}{{$\to$}}1 {^}{{$\mspace{-3mu}\widehat{\quad}\mspace{-3mu}$}}1
	{<}{$<$ }2 {>}{$>$ }2 {>=}{$\geq$ }2 {=<}{$\leq$ }2
	{<:}{{$<\mspace{-3mu}:$}}2 {:>}{{$:\mspace{-3mu}>$}}2
	{<=}{{$\Leftarrow$ }}2
	{=>}{{$\Rightarrow$ }}2 {+}{$+$ }2 {++}{{$+\mspace{-8mu}+$ }}2
	{<=>}{{$\Leftrightarrow$ }}2 {+}{$+$ }2 {++}{{$+\mspace{-8mu}+$ }}2
	{\~}{{$\mspace{-3mu}\widetilde{\quad}\mspace{-3mu}$}}1
	{!=}{$\neq$ }2 {*}{${}^{\ast}$}1 
	{\#}{$\#$}1
}
\newcommand{\setlistblue}[0]{\lstset{language=Java,numbersep=5pt,frame=single}}
\newcommand{\setlistred}[0]{\lstset{language=Ass,numbersep=5pt,frame=single}}

\newcommand{\mi}[1]{\ensuremath{\mathit{#1}}}

\newcommand{\mk}[1]{\mathfrak{#1}}
\newcommand{\mb}[1]{\mathbb{#1}}

\DeclareMathOperator{\ceq}{\ensuremath{{\simeq_{{ctx}}}}}
\DeclareMathOperator{\nceq}{{\not\simeq_{{ctx}}}}

\DeclareMathOperator{\treq}{{\overset{T}{=}}}

\DeclareMathOperator{\ntreqs}{\src{{\overset{T}{\neq}}}}

\DeclareMathOperator{\ntreqt}{\trg{{\overset{T}{\trgb{\neq}}}}}

\DeclareMathOperator{\ceqs}{\mathrel{\src{\ceq}}}
\DeclareMathOperator{\nceqs}{\mathrel{\src{\nceq}}}
\DeclareMathOperator{\ceqt}{\mathrel{\trg{\trgb{\simeq}_{{ctx}}}}}
\DeclareMathOperator{\nceqt}{\mathrel{\trg{\trgb{\not\simeq}_{ctx}}}}

\DeclareMathOperator{\linkingop}{\mathrel{\ltimes}}

\newcommand{\trsteptrg}[2][p]
{\ensuremath{\mathrel{\trg{ \xrightharpoonup{#2}_{#1}}}}}

\crefname{assumption}{assumption}{assumptions}
\Crefname{assumption}{Assumption}{Assumptions}
\crefname{fact}{fact}{facts}
\Crefname{fact}{Fact}{Facts}
\crefname{equation}{proposition}{propositions}
\Crefname{equation}{Proposition}{Propositions}
\crefname{example}{example}{examples}
\Crefname{example}{Example}{Examples}

\newcommand{\ctx}[0]{\ensuremath{\mk{C}}}
\newcommand{\ctxs}[0]{\src{\ctx}\xspace}
\newcommand{\ctxt}[0]{\trg{\ctx}\xspace}
\newcommand{\ctxh}[1]{\ctx\hole{#1}}

\newcommand{\hole}[1]{\ensuremath{\left[#1\right]}}

\newcommand{\cmp}[1]{\compilersymbol{#1}}

\newcommand{\tikzpic}[1]{
\begin{tikzpicture}[shorten >=1pt,auto,node distance=6mm]
\tikzstyle{state} =[fill=white,minimum size=4pt]
\tikzstyle{field} =[fill=gray!5,draw=black!70, rectangle, minimum width={width("whiskersfieldww")+2pt}]]
#1
\end{tikzpicture}
}

\usepackage{algorithmic}
\usepackage{graphicx}
\usepackage{textcomp}
\def\BibTeX{{\rm B\kern-.05em{\sc i\kern-.025em b}\kern-.08em
    T\kern-.1667em\lower.7ex\hbox{E}\kern-.125emX}}
\begin{document}

\title{CapablePtrs: Securely Compiling Partial Programs
	\\
	Using the Pointers-as-Capabilities Principle
}

\author{%
\IEEEauthorblockN{Akram El-Korashy}
\IEEEauthorblockA{MPI-SWS}
\and
\IEEEauthorblockN{Stelios Tsampas}
\IEEEauthorblockA{KU Leuven}
\and
\IEEEauthorblockN{Marco Patrignani}
\IEEEauthorblockA{CISPA}
\and
\IEEEauthorblockN{Dominique Devriese}
\IEEEauthorblockA{Vrije Universiteit Brussel}
\and
\IEEEauthorblockN{Deepak Garg}
\IEEEauthorblockA{MPI-SWS}
\and
\IEEEauthorblockN{Frank Piessens}
\IEEEauthorblockA{KU Leuven}
}


\maketitle

\begin{abstract}
Capability machines such as \cheri provide memory capabilities that
can be used by compilers to provide security benefits for compiled
code (e.g., memory safety).  The existing C to \cheri compiler, for
example, achieves memory safety by following a principle called
``pointers as capabilities'' (\pac). Informally, \pac says that a
compiler should represent a source language pointer as a machine code
capability.  But the security properties of \pac compilers are not yet
well understood. We show that memory safety is only one aspect, and
that \pac compilers can provide significant additional security
guarantees for {\em partial programs}: the compiler can provide
security guarantees for a compilation unit, even if that compilation
unit is later linked to attacker-provided machine code.

As such, this paper is the first to study the security of \pac
compilers for partial programs formally. We prove for a model of such
a compiler that it is {\em fully abstract}.  The proof uses a novel
proof technique (dubbed \tricl, read \emph{trickle}), which should be
of broad interest because it reuses the whole-program compiler
correctness relation for full abstraction, thus saving work.  We also
implement our scheme for C on \cheri, show that we can compile legacy
C code with minimal changes, and show that the performance overhead of
compiled code is roughly proportional to the number of
cross-compilation-unit function calls.

\end{abstract}


\numberwithin{lstlisting}{section}

\section{Introduction}\label{sec:intro}
In a conventional computer, memory is addressed using integers
(\emph{pointers}). In a capability machine, memory is addressed using
\emph{capabilities}~\cite{capbasedaddressing,hwcapbasedaddressing,saltzer1975protection,cheriageofrisk,cheriisatech,mmachine,capabook}, which carry  more
information than just a memory address---they also
contain \emph{bounds information}, indicating a range of memory that
can be accessed using the capability, and possibly also other
information such as access permissions. Load and store instructions
take a capability (in a register), and the machine checks that the
memory accessed is within the capability's bounds and that the
operation is compliant with the capability's permissions. If not, the
instruction fails with an exception.
The hardware also ensures that integers and capabilities are not
confused. 
One way of ensuring this
is by tracking capabilities in memory and in registers by \emph{tagging} memory locations and registers that contain capabilities.
 Hence, capability machines implement a more
structured memory model where (somewhat simplified) memory is a
collection of independent integer-indexed arrays containing integers
or capabilities, and every capability gives access to a contiguous
segment of one of those arrays.  This more structured memory model can
be used to implement \emph{fine-grained memory protection}, and has
the potential to provide protection against many software bugs.

A recent capability machine is \cheri~\cite{cheriageofrisk}. It has its own FreeBSD version
and C
compiler~\cite{watson2017balancing,cheriageofrisk,micro16,chericompartment,beyondpdp11}. Many
key design choices in \cheri were made to facilitate the use of memory
protection in existing, large code bases.  Specifically, \cheri
supports the pointers-as-capabilities (\pac) principle, which
intuitively dictates that a compiler should represent a
source-level \textit{pointer} as a target-level \textit{capability}.
To make this convenient, a \cheri capability contains (among other
things) $\mathtt{base}$ and $\mathtt{length}$ addresses, and an
$\mathtt{offset}$ relative to the base
address \cite{beyondpdp11}. Such a capability represents a pointer
pointing to the address $\mathtt{base} + \mathtt{offset}$, and that is
valid only if $(\mathtt{base} + \mathtt{offset}) \in
[\mathtt{base},\mathtt{base} + \mathtt{length})$.  Pointer arithmetic
can be implemented by manipulating the offset.
The following example illustrates how a compiler can map C pointers
to such machine-level capabilities.%
\footnote{
  As a typesetting convention, we use a \src{blue}, \src{sans\text{-}serif} font for \src{source} language elements and an \trg{orange}, \trg{bold} font for \trg{target} language ones.
  Elements common to both languages are typeset in a \com{\commoncol}, \com{italic} font.}



\begin{lstlisting}[mathescape]
extern void send_rcv(char* buffer);
static char iobuffer [512];
static int secret;

void f() {
	iobuffer[42] = 'X';
	send_rcv(iobuffer);
}
\end{lstlisting} 
\setlistred
\begin{lstlisting}[mathescape]
	csl $\$$c1, $\$$ddc, 512;
	li  $\$$r1, 'X';
	csw $\$$r1, 42($\$$c1);
	call send_rcv;
\end{lstlisting} 
\setlistblue
The C compilation unit (top) declares two module-scoped variables and defines a function \lstinline{f()} using one of these variables. 
	The assembly pseudocode (bottom) shows how a \pac compiler could translate the body of \lstinline{f()}.
	The {\em default data capability} register 
	\setlistred
	\lstinline{$ddc}
	\setlistblue contains a capability for the global data section.
	The compiler knows that the variable \lstinline{iobuffer} occupies the first 512 bytes of that global data section.
	Hence, the first instruction (\setlistred\lstinline{csl}\setlistblue, set length of a capability) loads in register 
	\setlistred
	\lstinline{$c1} a copy of \lstinline{$ddc}
	but with the \mbox{\lstinline{length} }
	\setlistblue
	field reduced to 512. The next two instructions implement the assignment instruction
	of \lstinline{f()}. Note that an out-of-bounds access would be trapped by the hardware. The final 
	\setlistred
	\lstinline{call}
	\setlistblue 
	instruction
	implements the function call in \lstinline{f()} (assuming a calling convention where the argument is passed in
	register 
	\setlistred\lstinline{$c1}\setlistblue). All accesses to \mbox{\lstinline{iobuffer}} performed in \lstinline{send_rcv} will be
	bounds-checked, since the capability passed to \lstinline{send_rcv} carries the bounds information.

\label{sec:intro-secprop}
Which security benefits does such a \pac compiler provide? First, the
compiler provides {\em spatial memory safety}. Since the bounds
meta-data for a pointer is stored together with the pointer address in
a single capability value, it is natural to implement a
bounds-checking compiler~\cite[Section 4.3]{watson2020cheri}. For
instance, an out-of-bounds access to \lstinline{iobuffer} in our
example will not access the \lstinline{secret} variable, but just
fail.\footnote{In principle, it is also possible to have the compiler
provide {\em temporal memory safety}, but this is harder as it
requires zeroing out all capabilities to a memory region when the
region is freed. Efficient temporal memory safety for C is still an
open problem and out of scope here. We refer the interested reader to
prior
work~\cite{10.1145/3434287,TsampasStelios2019Tsfs,skorstengaard2018reasoning,Skorstengaard:2019:SEW:3302515.3290332,xia2019cherivoke,filardo2020cornucopia}.}

However, this is not the full story regarding security properties.
Consider the example again under the assumption that the
external \lstinline{send_rcv} function is implemented directly in
assembly. Now, we lose the guarantee that \lstinline{send_rcv} cannot
access the \lstinline{secret} variable because an assembly level
implementation can directly access
\setlistred
\lstinline{$ddc}%
\setlistblue%
. Hence, upfront, memory safety is only guaranteed for {\em complete
programs}: if \emph{all} code in a program is compiled by the \pac compiler,
then all out-of-bounds accesses will be trapped.

Nonetheless, a \pac compiler can provide a security guarantee even
for \emph{partial programs} by relying on capability
unforgeability \cite{nienhuis2019rigorous} and on a trusted control
stack (that hides the capability of a program part from the context).
The main contribution of this paper is to prove that a \pac compiler
can, in fact, provide strong security guarantees for partial
programs. For our example, the compiler we model in this paper
provides the guarantee that \lstinline{secret} is inaccessible, even
if the function \lstinline{send_rcv} is implemented in hand-crafted
assembly.

To achieve non-trivial security guarantees for partial programs, the
target capability machine needs to support a mechanism to define
separate protection domains within a single process. For
instance, \cheri provides support for so-called \emph{object
capabilities} \cite{chericompartment,micro16}. This makes it possible
to put different program parts in separate protection domains.
\cheri mainly uses object capabilities to
{\em compartmentalize} programs: it offers an API to programmers to
run parts of a program in a {\em sandbox}, a protection domain with
reduced privileges. The current \cheri compiler, however, does not
make direct use of the object capability mechanism: it can only be
used through the provided API by the \emph{programmer} who has to
define and set up sandboxes manually. The
\pac compiler we propose in this paper 
\emph{automatically} sets
up a separate protection domain for every compilation unit. Doing so
allows the compiler to provide strong security guarantees for partial
programs.

\paragraph*{Summary of our results}
\label{sec:summary-of-results}
We study the security guarantees that a \pac compiler can provide
for \emph{partial programs}. Our setting is a \pac compiler from a
simple imperative source language with pointers to a capability
machine with memory capabilities and a very basic form of protection
domains/object capabilities. Our overarching contribution is a very
strong security theorem for this compiler, namely, \emph{full
abstraction
(\fa)}~\cite{abadiprotectiontranslations,Patrignani:2019:FAS:3303862.3280984},
which intuitively means that the compilation preserves and reflects
observational equivalence of partial programs. \fa implies the
preservation of many security properties like data confidentiality,
even when target contexts are arbitrary target code (in our case,
arbitrary assembly code) that may \emph{not} respect the compiler's
conventions.

In proving \fa for \pac, we make two additional contributions. Our
first contribution is a new trace-based proof technique for \fa that
can simultaneously handle \emph{dynamic memory sharing} between
modules and, importantly, \emph{reuses whole-program compiler
correctness as a black-box} to simplify the \fa proof. \fa proofs with
dynamic memory sharing are difficult and (whole-program) compiler
correctness is usually proved anyhow, so reusing it for proving \fa
reduces work. Technically, our proof is structured as a new 3-way
simulation called \tricl (read ``trickle'').  We expect \tricl to be
of interest beyond our \pac setting.

Second, to prove \fa, we find it essential to reflect some
structure of capabilities at the source-code level, forcing the
programmer to take into account some of the machine-code-level
expressiveness when reasoning about the source program. Interestingly,
not doing this may also lead to subtle security vulnerabilities as the
following example illustrates.

\begin{lstlisting}[mathescape]
static bool secret=true;

int branch_on_secret(int* p, int* q) {
	if ((int) p $!=$ (int) q) return 0;
	if (secret) return p[1]; else return q[1];
}	
\end{lstlisting}

Function \lstinline{branch_on_secret()} first tests whether its two
argument pointers are equal addresses. This equality implies that the
dereference operations \lstinline{p[1]} and
\lstinline{q[1]} both evaluate to the same value (or both fail).
Hence, \lstinline{branch_on_secret()} is not really meant to leak any
information about \lstinline{secret}.  However, a machine-code level
adversary can call \mbox{\lstinline{branch_on_secret()}} with two
capabilities that both point to the same address but that have
different bounds information. In that case, accessing \lstinline{p[1]}
could fail, while accessing \lstinline{q[1]} could succeed and return
a value.  Thus, the behavior of \mbox{\lstinline{branch_on_secret()}}
in that case {\em does} leak information
about \lstinline{secret}. More dangerously, it leaks this information
in a way to which the source-level programmer is oblivious. The
source-level programmer does not have a way to access bounds
information through pointer operations, while this bounds information
actually opens up an information channel.  Hence, to prove our \fa
result (which includes the preservation of source-level contextual
equivalence), we extend the source language to make pointers carry
bounds information. This essentially makes explicit exactly what
aspects of the target language programmers need to take into account
to reason about the security of partial source programs.

A second property we require for \fa is that machine code does not
have direct access to the program counter capability. This is easily
checked at link time. It guarantees that the target context cannot
confuse a partial program by providing it a code capability where it
expects a data capability (a behavior that does not exist in our
source language).

Finally, as our last contribution, we implement our compiler for
C by adding a ``compartmentalizing compiler-pass'' on top of the
existing C-to-\cheri compiler (the non-compartmentalizing \pac
compiler)~\cite{chericompartment,watson2020cheri}, and we evaluate the
performance cost as well as the compatibility of the
compartmentalizing \pac compiler with existing code.

\paragraph*{Contributions summary}
To summarize, we make the following contributions:
\begin{list}{$\bullet$}{\leftmargin=0em \itemindent=0.5em}
 \item We state and prove the security properties of a
pointers-as-capabilities (\pac) compiler for partial programs for the
first time. In doing so, we make several technical contributions:
 \begin{list}{$-$}{\leftmargin=1em \itemindent=1em}
  \item the definition of a sound and complete trace semantics for a
  C-like language (\Cref{sec:src-lang}) and for a language with
  capabilities (\Cref{sec:trg-lang}). Both languages feature a memory
  model that allows fine-grained dynamic memory sharing;
  \item the definition of a compiler between the aforementioned
  languages that embodies the \pac principle (\Cref{sec:compiler});
  \item a proof that the said compiler is fully abstract, with a new
  trace-based technique, \tricl, that handles dynamic memory sharing
  and allows reuse of whole-program compiler correctness for \fa
  (\Cref{sec:proofidea});
 \end{list}
 \item An implementation of our compiler on top of the existing
 C-to-\cheri compiler and a measurement of its efficiency and
 compatibility with existing C code
 (\Cref{sec:implementation,sec:experiments}).
\end{list}

We have simplified some aspects of the technical work for
presentation. The full details as well as proofs are provided in a
technical report (TR), available in a public repository~\cite{capable-ptrs-public}.
The implementation of our compiler and the related benchmarks are also
available in the same repository.



\section{Full abstraction and a new proof technique}
\label{sec:background}

We briefly recap compiler full abstraction (\fa), outline at a
high-level how it is proved, explain why a new proof technique is
needed and what our new technique (\tricl) does.


\subsection{Preliminaries}

Given execution states $s, s'$ of a language, and a small-step
reduction relation $\rightarrow$ (with a reflexive transitive closure
$\rightarrow^{*}$), we denote by $s \rightarrow s'$ the judgment that
state $s$ \emph{executes} and transitions into state $s'$. We call a
state \emph{stuck} if there is no $s'$ such that $s \rightarrow s'$. We treat
exceptions and silent divergence the same way we treat stuck states, so,
with slight abuse of terminology, we use the term \emph{diverging} for
executions that get stuck, silently reduce forever or end in
exceptions.

\subsubsection*{Programs, initial \& terminal states}

In our simple imperative setting, a (partial) program is a list of
modules, and a module is a list of functions. A program is
called \emph{whole} if its functions refer only to other functions
also defined within the program. Otherwise, it is called
partial. Linking of a pair of programs is denoted
$\mathrel{\ltimes}$. We define linking noncommutatively because this
simplifies some technicalities (see the TR for details). For the sake
of exposition, we distinguish two parts $\ctx$ and $p$ of a linked
program $\ctx \linkingop p$ as the \emph{context} and
the \emph{program}, suggesting that the latter is the \emph{program
part of interest} because it is the program that is translated by our
compiler.  Note that each of $\ctx$ and $p$ may themselves consist of
more than one \emph{module} (i.e., more than one compilation unit). As
usual, only whole programs with a \emph{main} entry-point function can
execute. We also use the standard notation $\ctx[p]$ for
$\ctx\linkingop p$.

The initial state of a program $p$ is denoted
$\mi{init}(p)$.
A state $s$ is called \emph{terminal} when it satisfies a special
judgment $\vdash_t s$.  If the execution of a program of interest $p$
in a certain context $\ctx$ reaches a terminal state, then we say the
execution \emph{converges} (or, that the
\emph{program} converges).
We denote convergence by $\ctxh{p} \Downarrow$,
which is shorthand for 
\(
\exists s.~\mi{init}(\ctx[p])
\rightarrow^{*} s ~\wedge~ \vdash_t s
\). Next, we define contextual equivalence of (partial) programs.


\begin{definition}[Contextual equivalence] 
	\label{def-ctxeq} \mbox{}\[p_1 \ceq
	p_2 \defeq \forall \ctx.~\ctxh{p_1} \Downarrow \iff \ctxh{p_2} \Downarrow\]
\end{definition}

\subsection{Compiler full abstraction}
\label{sec:faform}

A compiler \compilersymbol{\cdot} is fully abstract when it preserves
 and reflects contextual equivalence. The use of full abstraction to
 establish compiler security is
 standard~\cite{abadiprotectiontranslations,fullyabstractjavascript,Patrignani:2019:FAS:3303862.3280984}.
\begin{definition}[Full abstraction]\label{theorem-full-abstraction}
The compiler $\compilersymbol{\cdot}$ is \fa if for all $\src{p_{s1}}, \src{p_{s2}}$:
\begin{enumerate}[label=(\roman*)]
    \item \label{condition-reflects}
    (Reflection)
    $\compilersymbol{\src{p_{s1}}} ~\ceqt~
    \compilersymbol{\src{p_{s2}}} \implies 
    \src{p_{s1} ~\ceqs~ p_{s2}}$
    \item \label{condition-preserves}
    (Preservation)
    $\compilersymbol{\src{p_{s1}}} ~\ceqt~
    \compilersymbol{\src{p_{s2}}} \impliedby 
    \src{p_{s1} ~\ceqs~ p_{s2}}$
\end{enumerate}
\end{definition}

Condition \ref{condition-reflects} ensures that the compiler is
non-trivial (a trivial compiler might compile semantically different
programs to the same output program, which is forbidden by
reflection).  Reflection usually follows immediately
from \emph{backward simulation}, the standard formalization of the
compiler's whole-program
correctness \cite{leroy2009formal,patterson2019next}.

Condition \ref{condition-preserves} is the ``security-relevant''
direction of \fa as it ensures that no extra distinguishing power is
gained by target contexts as compared to source contexts. It implies
the preservation of any security property that can be formalized as
program equivalence (a notable example being noninterference for
confidentiality).
It is usually proved in the contrapositive: Assume
$\compilersymbol{\src{p_{s1}}}
~\nceqt~ \compilersymbol{\src{p_{s2}}}$, and show $\src{p_{s1}
~\nceqs~ p_{s2}}$. From the assumption, there must be a target context
that distinguishes $\compilersymbol{\src{p_{s1}}}$ and
$\compilersymbol{\src{p_{s2}}}$ (causes one to diverge and the other
to converge). From this, we need to construct a source context that
distinguishes $\src{p_{s1}}$ and $\src{p_{s2}}$. This construction of
the source context is called \emph{back-translation} in the
literature. There are two broad approaches to
back-translation: \emph{syntax-directed}
and \emph{trace-directed}~\cite{Patrignani:2019:FAS:3303862.3280984}. Here,
we follow the trace-directed approach as we find it technically more
convenient (we discuss syntax-directed approaches
in \Cref{sec:related-work}).

The key idea of the trace-directed approach is to characterize
the \emph{observable} behavior of partial programs via
a \emph{labeled} transition system that produces \emph{(finite)
traces} describing how a partial program interacts with its
environment.\footnote{For the purpose of proving \fa, finite traces
suffice, so ``trace'' in this paper always refers to a finite trace.}
This is done separately for source and target languages. We denote the
set of traces of a partial program $p$ by $Tr(p)$. Next,
define \emph{trace equivalence} $\treq$ of partial programs in source
and target languages separately: Two partial programs (both source or
both target) $p_1, p_2$ are trace equivalent, $p_1 \treq p_2$, if
$Tr(p_1) = Tr(p_2)$. We then prove three lemmas.
\begin{lemma}[Soundness of target trace equivalence]
	\label{soundness-target-trace-equivalence}
	$
	\compilersymbol{\src{p_{s1}}} ~\nceqt~
	\compilersymbol{\src{p_{s2}}} \implies
	\compilersymbol{\src{p_{s1}}} ~\ntreqt~
	\compilersymbol{\src{p_{s2}}}
	$
\end{lemma}

\begin{lemma}[Compilation preserves trace equivalence]
	\label{compilation-preserves-trace-equivalence}
	$
	\compilersymbol{\src{p_{s1}}} ~\ntreqt~
	\compilersymbol{\src{p_{s2}}} \implies
	\src{p_{s1}} ~\ntreqs~
	\src{p_{s2}}
	$
\end{lemma}

\begin{lemma}[Completeness of source trace-equivalence]
	\label{completeness-source-trace-equivalence}
	$
	\src{p_{s1}} ~\ntreqs~
	\src{p_{s2}} \implies
	\src{p_{s1}} ~\nceqs~
	\src{p_{s2}}
	$
\end{lemma}

The composition of these three lemmas immediately yields our goal, the
contrapositive of condition~\ref{condition-preserves}! The important
point is that only Lemma~\ref{compilation-preserves-trace-equivalence}
bridges the two languages\footnote{Although
	\Cref{soundness-target-trace-equivalence} mentions
the compiler (i.e., appears to also bridge the two
languages), we found that its proof does not rely on
a cross-language simulation, which is what we really
mean by bridging the two languages.
}.
%
%
Lemma~\ref{compilation-preserves-trace-equivalence} follows
immediately from the following two lemmas, which together say that the
compiler preserves and reflects traces of partial programs.
\begin{lemma}[No trace is omitted by compilation]
	\label{no-trace-omitted}
	$\alpha \in \src{{Tr}(p_s)}
	\implies \alpha \in \trg{{Tr}(}\compilersymbol{\src{p_s}}\trg{)}$
\end{lemma}
\begin{lemma}[No trace is added by compilation]
	\label{no-trace-added}
	$\alpha \in \src{{Tr}(p_s)}
	\impliedby \alpha \in \trg{Tr(}\compilersymbol{\src{p_s}}\trg{)}$
\end{lemma}

Lemma~\ref{no-trace-omitted} follows directly from compiler (forward)
simulation, which is needed to prove the compiler correct
anyhow. Hence, this lemma does not add additional proof effort.
On the other hand, Lemma~\ref{no-trace-added} is an additional (and
the last remaining!) proof burden. The ``difficulty'' of this proof
really depends on the complexity of the traces, i.e., the complexity
of interaction between a program and its context.


\subsection{The Why and What of \tricl}

In prior work that uses trace-based back-translation, program modules
cannot share memory or the shared memory is fixed
upfront~\cite{Abate:2018:GCG:3243734.3243745,patrignani2015secure,Patrignani:2019:FAS:3303862.3280984}. Traces are easy to
define in this setting. In our \pac setting, however, the memory
shared between modules can grow dynamically as the program shares more
of its previously private memory with the context by passing it
corresponding capabilities (or pointers in the source). The context
can also pass capabilities back to the program but it \emph{should}
only pass back those capabilities that it started with or those that
it received previously during the execution (since capabilities cannot
be forged by design). Consequently, any trace on which the context
passes back a capability that it didn't receive earlier should be
removed from consideration as invalid. However, this notion of trace
(in)validity is not straightforward (in either language) because
capabilities can be passed not just directly via function arguments,
but also \emph{indirectly}, as something that's reachable in the heap
at the point of a transition. This makes valid traces
on \emph{partial} programs rather hard to define. To avoid this
problem, we use a different definition of traces.

\paragraph*{A new definition of traces}
We start from \emph{whole} programs, not partial programs. The trace
is defined by decorating the small-step semantics of the whole program
with information about interaction between its modules. We then define
the traces of a partial program as the traces it can produce when
linked with \emph{some (existentially quantified) context}. The
explicit presence of the context automatically ensures that all traces
are valid, so we don't have to define validity separately. (In other
words, validity of a trace now follows implicitly from
easier-to-define invariants on the whole execution state.)

\begin{definition}[Traces of a partial program]
    \label{def-tr}
\[ \mi{Tr}(p) \defeq \{ \alpha \mid \exists \ctx, s, \varsigma.\, (\mi{init}(\ctx[p]), \emptyset) \paronto{\neutralcol{\alpha}}_{p} (s, \varsigma) \} \]
\end{definition}

$\emptyset$ and $\varsigma$ in this definition are auxiliary state
components, that we describe in \Cref{sec:proofidea}. The reader can
ignore them for now. $\paronto{\neutralcol{\alpha}}_{p}$ represents
label-decorated whole-program reduction, which we also define
in \Cref{sec:proofidea}.  To prove Lemma~\ref{no-trace-added} with
this definition of traces, we must show that, given a trace $\alpha$
of $\compilersymbol{\src{p_s}}$, there is an \emph{emulating} source
context $\src{\ctx_{emu}}$ such that $\src{\ctx_{emu}[p_s]}$ produces
$\alpha$. In other words, we must \emph{back-translate} a trace
$\alpha$ into a source context $\src{\ctx_{emu}}$ and show that
$\src{\ctx_{emu}[p_s]}$ can actually produce $\alpha$. For this, we
would like to set up a simulation between the target and source
runs. However, this simulation can be very hard to set up because we
constructed the emulating source context without considering the given
target context (just from the trace prefix) and, hence, there can be
differences between the \emph{internal} behaviors of the emulating
context and the given target context, e.g., in the specific order of
updates to the shared memory, and in the internal function
calls. These differences mean that the memory (and the call stack) do
not remain in sync between the emulating source context and the given
target context and our simulation relation needs to accommodate the
big gap between the internal states of the target contexts and the
emulating source contexts. We call this the ``vertical gap''.

\paragraph*{\tricl}
This is where our new \tricl technique comes in. \tricl introduces a
third ``mediator'' run to the simulation, namely, that induced by the
\emph{compilation of the whole source program} containing both
\src{p_s} and \src{\ctxs_{emu}}. Overall,
the three runs are (i) the given (target) run of
\trg{\ctxt[\cmp{\src{p_s}}]}, where \trg{\ctxt} is the target context
that induces the trace $\alpha$, (ii) the emulating source run
of \src{\ctx_{emu}[p_s]}, and (iii) the mediating target run of
\cmp{\src{\ctxs_{emu}[p_s]}}.

Introducing the mediator run simplifies the proof as follows. Because
the mediator \cmp{\src{\ctxs_{emu}[p_s]}} is obtained by a whole
program compilation of the emulator \src{\ctx_{emu}[p_s]}, we
can \emph{reuse whole-program compiler correctness as a black box} to
immediately reduce the problem of showing that the emulating (source)
run emulates the given (target) run to that of showing that the
mediating (target) run emulates the given (target) run. This reduced
problem is now about \emph{two runs in the same language} (target),
which enables us to write a big part of our simulation invariants as
same-language (instead of cross-language) invariants.  In other words,
this simplifies the ``vertical gap''.
 
\Cref{sec:proofidea} explains the \tricl simulation for our \pac
compiler, but we note that the idea is general and should apply to
other settings as well.

\section{Source and target languages}
\label{sec:languages}
Next, we introduce our source and
target languages, \sourcelanguage (\Cref{sec:src-lang})
and \targetlanguage (\Cref{sec:trg-lang}).


\subsection{\sourcelanguage: The source language}
\label{sec:src-lang}
To keep our focus limited to the ``pointers as capabilities'' aspect
of the translation, we design \sourcelanguage and \targetlanguage to
be pretty close except in how they deal with memory. For
example, \sourcelanguage features only unstructured control flow in
the form of a
\src{\ms{JumpIfZero}} instruction.
But it still has functions and modules.  In fact, \emph{modules} are
 crucial.  They are units of memory isolation, which makes programs
 interesting for security. Briefly,
\begin{enumerate*}
\item every module gets its own \emph{module-global
variables}; every function inside the module can access these
variables, whereas any function external to the module cannot access
these variables directly by default, and
\item every module gets its own \emph{data
stack} on which it stores the frames of live (ongoing) calls to its
functions.
\end{enumerate*}

\begin{figure}
\begin{lstlisting}[
caption= \sourcelanguage module. 
,
frame=tlrb,label=src-lang-example,
escapechar=|]
Module id: Main
Import module: Networking

iobuffer [512]; |\label{line:iobuffer_decl}|
secret; |\label{line:secret_decl}|

main() {
  Assign &iobuffer[42] 4242; |\label{line:iobuffer_ass}|
  Call read_secret(); |\label{line:call_read_secret}|
  Call encrypt(); |\label{line:call_encrypt}|
  Call send_rcv(&iobuffer); |\label{line:callsend_rcv}|
  Call decrypt();
}

read_secret() { ... }
encrypt() { ... }
decrypt() { ... }
\end{lstlisting}
\end{figure}

For example, \Cref{src-lang-example},
shows a module (\lstinline|Main|)
with two module-global variables, \lstinline{iobuffer}
and \lstinline{secret} (lines \ref{line:iobuffer_decl} \& 
\ref{line:secret_decl}).
The \lstinline{main} function\footnote{Notice
in the \lstinline|main| function on line \ref{line:iobuffer_ass}
the use of
variable \lstinline|iobuffer|;
the syntax for l-values is more explicit
than in C.
} and the secret-handling
functions \mbox{\lstinline|read_secret|}, \lstinline|encrypt| and
\lstinline|decrypt|    
are defined in the same module, and thus can access the
variables \lstinline|iobuffer| and \mbox{\lstinline|secret|}.  The
function \lstinline{send_rcv}, in contrast, is external: it is defined
in the \mbox{\lstinline{Networking}} module which is presumably untrusted.
On line \ref{line:callsend_rcv}, the \lstinline|Networking|
module gains access to
\lstinline|iobuffer|.
The \mbox{\lstinline|Networking|} module, however, does not gain access
to \mbox{\lstinline|secret|} through \mbox{\lstinline|&iobuffer|}---so
 long as the other trusted functions \lstinline|read_secret|
and \lstinline|encrypt| also make sure to
 not copy (pointers to) \mbox{\lstinline|secret|} into the \lstinline|iobuffer|.
Any attempt to increment and access the pointer
 \lstinline|&iobuffer| beyond the array bounds gets stuck.
To understand how we model these bounds checks, we introduce 
the expression semantics and the memory model of
\sourcelanguage.

\subsubsection{Expressions and memory model of
     \sourcelanguage}
\label{sec:expr-memory-model}

Expressions are denoted \src{e} and do not update memory.
\begin{align*}
\src{e}  ::=\   & \src{\mathbb{Z}} \mid \src{ VarID} \mid \src{e \oplus e} \mid \src{e[e]} \mid \src{\&VarID} \mid \src{\&e[e]} \mid ~\src{^{*}(e)} \mid
\src{start(e)} \\ & \mid \src{end(e)} \mid \src{offset(e)} \mid \src{capType(e)} \mid \src{limRange(e, e, e)} \\
\src{{V}} ::=\  & \src{\mb{Z}} \mid \src{{Cap}}
\end{align*}
\emph{Base expressions} are integers \src{\mb{Z}} and
 variable identifiers
\src{VarID} (both local and global).
\emph{Binary arithmetic expressions}
 are generically
denoted \src{e \oplus e}. 
\emph{Pointer and array expressions}
are: 
\begin{enumerate*}
\item the array-offset expression \src{e[e]},
\item the ampersand (address-of) operator of the forms \src{\& {VarID}} and
\src{\& e[e]}, and
\item the star (pointer dereference) operator \src{^{*}(e)}
\end{enumerate*}. The intuitive meanings of these expressions are as in C.
Moreover, there are \emph{low-level expressions} that are necessary for
reflecting the target memory model in the source language (as mentioned in \Cref{sec:intro-secprop}):
four getters: \src{{start}(e)}, \src{{end}(e)},
\src{{offset}(e)}, and \src{{capType}(e)}; and a setter:
\src{{limRange}(e, e, e)}.
These low-level expressions operate on a capability-based 
representation
of memory addresses that we explain next.

Addresses in \sourcelanguage are represented as capabilities.
Thus, built into the memory model is a type
\src{{Cap}} for capabilities that is distinct from integers \src{\mb{Z}}.
Hence, run-time values \src{{V}} are integers \src{\mb{Z}} or
 capabilities \src{{Cap}}. A
 memory \src{{Mem}: \mb{Z} \xrightharpoonup{\ms{fin}}{V}} is a finite
 map from addresses to values. Note that the range of a memory may
 contain capability values.  Concretely, we define the type \src{Cap}
 of capability values
 as \src{{Cap} \defeq \{\kappa, \delta \} \times \mb{Z} \times \mb{Z} \times \mb{Z}}.
 The first field of the capability value indicates its type: code
 (\src{\kappa}) or data (\src{\delta}). The getter of this field is
 the expression \src{\ms{capType}}.  The next three (integer) fields
 are respectively the \src{\ms{start}},
\src{\ms{end}}, and \src{\ms{offset}} of the capability.
The start and end identify the memory region on which this capability
authorizes a code (\src{\kappa}) or data (\src{\delta}) access
operation.  The offset designates the address at which an access
operation is performed. The offset should be within range (checked
by \Cref{tr:evaluate-expr-deref} below).

The semantics of expressions are defined in \Cref{fig:expr-source} by
the judgment $\src{e \Downarrow v}$.  Notice that expressions in
\sourcelanguage~do not have side effects.
The evaluation context for expressions consists of:
\begin{enumerate}
	\item \textbf{syntactic} information about the program given by
	the function definitions \src{Fd}, the declarations of module-global
	variables \src{MVar}, and the layout and bounds \src{\beta} of
	all the program's variables;
	
	\item \textbf{load-time} information about the program given by
	the per-module \textbf{d}ata-segment-location \src{\Delta}, and the 
	per-module local-\textbf{s}tack-location
	\src{\Sigma}; and
	
	\item \textbf{execution-state} information given by the memory
	\src{Mem}, the stack pointers \src{\Phi}
	of the module-local stacks and the
	program counter \src{pc}.
\end{enumerate}

We make two observations on the rules of \Cref{fig:expr-source}.
First, \Cref{tr:evaluate-expr-addr-local,tr:evaluate-expr-addr-module}
are the only two rules (of expression evaluation) where a capability
value \emph{originates}. In fact, this capability value originates out
of thin air. This behavior (the minting of a capability value during
the course of the program execution) is precisely why this source
semantics is not really executable as is on a capability
architecture. On a capability architecture, a program does not have
any means to forge a capability value. Instead, it can only refer to
capability registers\footnote{It can also get capabilities by calling
the memory allocator, which in turn will place in a register (or in
the program's memory) a new capability authorizing access to the
allocated region.}.  (Contrast these two rules
to \Cref{tr:evalddc,tr:evalstc} of \Cref{sec:trg-lang}.)  Replacing
the mentions of program variables with expressions that mention
capability registers is thus the key role of a \pac compiler.

Second, \Cref{tr:evaluate-expr-deref} performs a bounds check before
it loads a value from memory. Building-in this check is how we
define \sourcelanguage~to be spatially safe by relying on the
capability-based memory model. A similar bounds check is performed
at \emph{store} time as well (see the rule for the \src{Assign}
command in the TR).

\begin{figure*}
	\centering
	\typerule{Eval-amp-local-var}{
		\src{({fid}, \_) = {pc}} &
		\src{{vid} \in \fun{localIDs}{\src{Fd(fid)}}
			\cup \fun{args}{\src{Fd(fid)}}} &
		\src{{mid} = \fun{moduleID}{\src{Fd(fid)}}} & \\
		\src{{\beta(vid, fid,  mid)} = [{st},{end})} &
		\src{\phi = \Sigma({mid}).1 + \Phi({mid})}
	}
	{\src{\&{vid} \Downarrow
			(\delta, {\phi} + {st}, {\phi} + {end}, 0)}
	}{evaluate-expr-addr-local}
	\typerule{Eval-amp-module-var}{
		\src{({fid}, \_) = {pc}} &
		\src{{vid} \notin \fun{localIDs}{\src{Fd(fid)}}
			\cup \fun{args}{\src{Fd(fid)}}} & 
		\src{{mid} = \fun{moduleID}{\src{Fd(fid)}}} & \\
		\src{{vid} \in {MVar}({mid})} &
		\src{{\beta(vid, \bot,  mid)} = [{st},{end})}
	}{\src{\&{vid} \Downarrow 
			(\delta, \Delta({mid}).1 + {st}, \Delta({mid}).1 +
			{end}, 0)}
	}{evaluate-expr-addr-module}
	\typerule{Eval-amp-arr}{
		\src{\&e_{arr} \Downarrow 
			(\delta, {st}, {end}, {off})} & 
		\src{e_{idx} \Downarrow {off'}} & 
		\src{{off'} \in \mathbb{Z}
		}
	}{\src{\&{e_{arr}[e_{idx}]} 
			\Downarrow (\delta, {st}, {end}, {off + off'})}
	}{evaluate-expr-addr-arr}
	\typerule{Eval-star}{
		\src{e \Downarrow (\delta, {st}, {end}, {off})} & 
		\src{{st} \leq {st} + {off} < {end}}
	}{\src{^{*}(e) \Downarrow {Mem}({st} + {off})}
	}{evaluate-expr-deref}
	\caption{(Excerpt) Evaluation of expressions
		in \sourcelanguage.  The evaluation
		relation \src{\Downarrow} is indexed with an
		evaluation context \src{{Fd},
		{MVar}, \beta, \Delta, \Sigma, {Mem}, \Phi, {pc}},
		which we elide for brevity.}  \label{fig:expr-source}
\end{figure*}


\subsubsection{Commands and execution state
     of \sourcelanguage}
 
Commands of \sourcelanguage are denoted \src{Cmd}. Unlike
expressions, they can modify memory and other parts of the execution state.
\begin{align*}
\src{Cmd} ::= ~&\src{Assign~e_l~e_r} \mid \src{Alloc~e_l~e_{size}} \mid
\src{Call~fid~\OB{e}} \mid \src{Return}
\mid 
\\ &~
\src{JumpIfZero~e_c~e_{off}} 
\mid
\src{Exit}
\end{align*}
The commands should be self-explanatory (\src{fid} is a function
name).  An execution state $\src{s \defeq \langle
Mem, \Phi, pc, stk, nalloc \rangle}$ of a program
in \sourcelanguage consists of the memory \src{Mem}, the stack
pointers \src{\Phi} of the module-local stacks, the program
counter \src{pc}, a trusted \emph{control} stack
\src{stk}---which is shared among all
 modules of a program---and
the memory-allocation status represented by the 
\emph{next-free-address} \src{nalloc}.
The space for dynamic memory allocation (i.e., the heap)
 is also shared by all the program modules  (like the trusted control stack).
The purpose of modeling a control stack that is trusted and 
hence separate from the accessible memory
is that we want to rule out all ill-formed control sequences.
No command in \sourcelanguage can write directly to this control stack.

The semantics are small step and use the
judgment \src{\rightarrow~ \subseteq s \times s} that is indexed with
an evaluation context \src{{Fd}, {MVar}, \beta, \Delta, \Sigma} and an
allocation limit \src{\negativeInfinity}.  The allocation limit
(\src{\negativeInfinity}) is the maximum possible size of the shared
heap. See the TR for details of the semantics.

\subsection{\targetlanguage: The target language}
\label{sec:trg-lang}
Our target language \targetlanguage is, like \sourcelanguage, an
imperative language with modules. However, unlike \sourcelanguage, it
does not feature variables. Instead, it only features ``capability
registers'' and integers as base expressions.  The role of the
compiler from \sourcelanguage to \targetlanguage is to implement
operations on source pointers by using operations on capability
registers.  The memory model is like that of \sourcelanguage
(\Cref{sec:expr-memory-model}), i.e., it is capability based.  Through
this capability-based memory model and capability
registers, \targetlanguage models a slightly abstracted and simplified
version of CHERI assembly~\cite{cheriisatech}.

\subsubsection{Expressions and commands in \targetlanguage}

\begin{align*}
\trg{e} ::=\ \trg{\mb{Z}} &\mid \trg{getddc} \mid \trg{getstc}
\mid \trg{e \oplus e} \mid \trg{inc(e, e)} \mid \trg{deref(e)}
\\
&
\mid \trg{start(e)}
\mid \trg{end(e)} \mid \trg{offset(e)} \mid
\trg{capType(e)}
\\
&
\mid \trg{limRange(e, e, e)}
\end{align*}
Expressions are denoted \trg{e}.
The language has three named capability
registers: \trg{stc}, \trg{ddc} and \trg{pcc}.  The names of these
registers hint at their recommended usage: \trg{ddc} stands for
default data capability while \trg{stc} stands for stack
capability. Our compiler uses \trg{ddc} as a capability on the
per-module data segment, and \trg{stc} on the per-module stack. The
third capability register, \trg{pcc}, holds the program counter
capability which points to the current command and allows the
execution of commands.
The expressions \trg{getddc} and \trg{getstc} immediately
return the current value of the respective capability
registers, \trg{ddc} and \trg{stc}.  (See rule \ref{tr:evalddc}
in \Cref{fig:expr-trg}.)  However, \trg{pcc} cannot be read
in \targetlanguage. This is a simple way of enforcing the restriction
(mentioned in \Cref{sec:intro-secprop}) on linking with contexts that
mention the \trg{pcc} register. No expression allows the fabrication
of an arbitrary capability, thus enforcing \emph{capability
unforgeability}.

\begin{figure*}
	\centering
	\typerule{Eval-ddc}{
    }{\trg{Mem, {ddc}, {stc}, {pcc} \vdash getddc \Downarrow
        {ddc}}
	}{evalddc}
	\typerule{Eval-stc}{
}{\trg{Mem, {ddc}, {stc}, {pcc} \vdash getstc \Downarrow
		{stc}}
}{evalstc}
\typerule{Eval-inc}{
	\trg{Mem, {ddc}, {stc}, {pcc} \vdash e \Downarrow
		(x, st, end, off)} &
	\trg{Mem, {ddc}, {stc}, {pcc} \vdash e_z \Downarrow
		v_z} &
	\trg{v_z \in \mb{Z}}
}{\trg{Mem, {ddc}, {stc}, {pcc} \vdash inc(e, e_z) \Downarrow
		(x, st, end, off + v_z)}
}{evalinc}	
\typerule{Eval-deref}{
	\trg{Mem, {ddc}, {stc}, {pcc} \vdash e \Downarrow
		(\delta, st, end, off)} &
	\trg{st \leq st + off < end}
}{\trg{Mem, {ddc}, {stc}, {pcc} \vdash deref(e) \Downarrow 
		Mem(st + off)
}}{evalderef}
\typerule{Eval-limRange}{
	\trg{Mem, {ddc}, {stc}, {pcc} \vdash e_1 \Downarrow
		(x, st, end, off)} & 
	\trg{Mem, {ddc}, {stc}, {pcc} \vdash e_2 \Downarrow
		st'} & \\
	\trg{Mem, {ddc}, {stc}, {pcc} \vdash e_3 \Downarrow
		end'} & 
	\trg{st' \in \mb{Z}} &
	\trg{end' \in \mb{Z}} &
	\trg{[st', end') \subseteq [st, end)} &
}{\trg{Mem, {ddc}, {stc}, {pcc} \vdash limRange(e_1, e_2, e_3) \Downarrow 
		(x, st', end', 0)
}}{evallimrange}
\caption{(Excerpt) Evaluation of expressions in \targetlanguage.}
\label{fig:expr-trg}
\end{figure*}

In \targetlanguage, one can 
increment (decrement) the offset of a capability by an arbitrary integer value (see
\cref{tr:evalinc}),
resulting in a new capability that nevertheless has the same bounds as
the original.  The check that the offset lies within bounds is only
performed at use time (e.g., use by means of a \trg{deref}
expression).  Observe that \cref{tr:evalderef} performs the same
bounds-check on the capability value
that \cref{tr:evaluate-expr-deref} (of \sourcelanguage)
performs. \trg{limRange(e_1, e_2, e_3)} \emph{restricts} the lower-
and upper-bounds of the capability \trg{e_1} to the
interval \trg{[e_2, e_3)}, returning a new capability.

Commands in \targetlanguage are the same as commands
in \sourcelanguage (modulo expressions). 
Briefly, the small-step command reduction is denoted
with \trg{\rightarrow~ \subseteq s \times s}
where a state \trg{s} of a \targetlanguage~program, as in \sourcelanguage, 
consists of a memory
\trg{Mem}, an allocation status \trg{nalloc} and a trusted stack
\trg{stk}, but, unlike \sourcelanguage, additionally
consists of three capability registers:
\trg{ddc}, \trg{stc} and \trg{pcc}, and a map \trg{mstc} holding a per-module capability
authorizing access to the module's local \emph{data} stack
(the local data stack is part of the memory \trg{Mem}).

Note that we model both a trusted stack (and hence a secure calling
convention), and a separate map for the per-module data-stack
capability in \targetlanguage.  This built-in segregation may sound
too abstract for a \emph{target} language that has low-level
elements like capability registers.  However, this modeling choice
allows us to focus only on the \pac principle for program variables
without worrying about the integrity of the stack pointer.  (Prior work has
already shown how compilers can enforce well-bracketed control flow
and stack encapsulation using
capabilities \cite{Skorstengaard:2019:SEW:3302515.3290332}.)


%

\section{Our \pac compiler}
\label{sec:compiler}

Our \sourcelanguage to \targetlanguage compiler
(\compilersymbol{\cdot}) translates pointers to capabilities (the \pac
principle).  In this section, we present its crucial bits, namely, the
translation of \sourcelanguage expressions to \targetlanguage
expressions.  The translations of commands
(denoted \instructiontrans{\cdot}) and of modules
(\compilersymbol{\cdot}) are rather trivial due to the similarity of
the syntax of the source and target commands and modules, so we elide
those.

The translation of expressions 
$\expressiontrans{\cdot}: \src{e} \rightarrow \trg{e}$, whose excerpts are presented below,
is indexed by the syntactic information \src{fid, modID} (function id
and module id) providing the scope of the expression being translated,
and \src{\beta}, giving the layout and bounds of source variables.
    \begin{align*}
    \expressiontrans{\src{z}}_\_ &~\defeq~\trg{z} & 
    \\
    \expressiontrans{\src{e_1 \oplus e_2}}_{~\src{fid, modID, \beta}}
    &~\defeq~
    \expressiontrans{\src{e_1}}_{~\src{fid, modID, \beta}}
    ~\trg{\oplus}~
    \expressiontrans{\src{e_2}}_{~\src{fid, modID, \beta}}&
    \\
    \expressiontrans{
        \src{\&vid}
    }_{
        ~\src{\_, modID, \beta}
    }
        &~\defeq~
            \trg{
                \begin{aligned}[t]
                 &
                \trg{limRange(getddc,}
                 \\
                 &\
                \trg{~start(getddc) + \src{st},}\\ &\
                \trg{~start(getddc) + \src{end})}
                \end{aligned}
            } 
    \\
    &\text{when}~ \src{\beta(vid, \bot, modID) = (st, end)} 
    \\
    \expressiontrans{\src{\&vid}}_{~\src{fid,
            modID, \beta}}
                &~\defeq~
                \trg{
                    \begin{aligned}[t]
                        &\neutralcol{\text{let}~s = \trg{getstc}~\text{in}}\\
                        &\neutralcol{\text{let}~\textit{so} = 
                        \trg{start(\neutralcol{s}) + offset(\neutralcol{s})}~\text{in}}\\
                        &\trg{limRange(\neutralcol{s},~}
                        \trg{\src{st} + \neutralcol{\textit{so}},~}
                        \trg{\src{end} + \neutralcol{\textit{so}})}
                	\end{aligned}        
                    }
    \\ &
    \text{when}~\src{\beta(vid, fid, modID) = (st, end)}
    \\
    \expressiontrans{\src{vid}}_{~\src{
            fid, mid, \beta}}
    &~\defeq~ \trg{deref(}
    \expressiontrans{\src{\&vid}}_{\src{
            fid, mid, \beta}}\trg{)}& \\
    \expressiontrans{\src{\&e_{arr}[e_{off}]}}_{~\src{fid, mid, \beta}}
    &~\defeq~ \trg{inc(}
    \expressiontrans{\src{\&e_{arr}}}_{~\src{fid, mid, \beta}},
    \expressiontrans{\src{e_{off}}}_{~\src{fid, mid, \beta}}\trg{)} &\\
    \expressiontrans{\src{e_{arr}[e_{off}]}}_{~\src{fid, mid, \beta}}
    &~\defeq~
    \trg{deref(} 
    \expressiontrans{\src{\&e_{arr}[e_{off}]}}_{~\src{fid, mid, \beta}}
    \trg{)} &
    \end{align*}
Translating expressions \src{start}, \src{end}, 
\src{offset}, \src{capType}, and \src{limRange} is
straightforward and is similar to
\expressiontrans{\src{e_1 \oplus e_2}}.
As an example, we show the translation of \Cref{line:callsend_rcv}
of \Cref{src-lang-example}.
\begin{align*}
&\instructiontrans{\src{Call~send\_rcv(\&iobuffer)}}
\\
=\ 
&
\trg{Call~send\_rcv~(limRange(}
\\
& \quad 
\trg{ddc, start(ddc) + 0, start(ddc) + 512))}
\end{align*}

Note that the compiler uses the bounds information that is given in
the text of the program in the declaration of the array
(\Cref{line:iobuffer_decl}) to introduce explicit curbing
(using \trg{limRange}) of the \trg{ddc} capability so that the
resulting capability is of the same size as the declared array size
(\trg{512}) and not bigger.  This curbing prevents the external
function \lstinline|send_rcv| from offsetting the capability beyond
the span of \lstinline|iobuffer| and from accessing other variables
like \lstinline|secret|. This curbing is, in fact, essential for full
abstraction since a similar out-of-bounds access is prohibited in the
source semantics.

\section{Proving the compiler fully abstract}
\label{sec:proofidea}

We prove our compiler fully abstract following the steps in
\Cref{sec:background}. We fill in details of two of the steps here:
The definition of traces and the proof of Lemma~\ref{no-trace-added}
using \tricl.

\subsection{Traces}\label{sec:treq}

As explained in \Cref{sec:background}, we define traces by augmenting
the small-step semantics of whole programs (\Cref{sec:languages}) with
labels that capture information about the interaction between the
program $p$ of interest and its context.  In our two languages, such
an interaction happens only through shared memory or function
arguments in only those steps that transfer control from the program
to the context or vice-versa. Accordingly, a trace label arises only
at such \emph{border-crossing} control transfer steps, and the label
records the shared memory, and the function arguments as
in~\citet{9249fdc52dd3414e83b2e3f9a89cb117}.  In the following, we
give a formal account of this development for the target language
\targetlanguage; the account for the source language \sourcelanguage
is similar.

The labeled trace-step relation for \targetlanguage is written
\trsteptrg[p]{\neutralcol{\lambda}}. The relation relates two
\emph{trace states} and a label $\lambda$.
A trace state \trg{(s, \neutralcol{\varsigma})} extends the normal
execution state \trg{s} with auxiliary information $\varsigma$, which
is the set of memory addresses shared so far (i.e., from the initial
state and up until execution state \trg{s}) between the program of
interest \trg{p} and the context.  This auxiliary information is used
to define an informative trace label $\lambda$. A trace $\alpha$ is a
finite list $\OB{\lambda}$ of labels $\lambda$.  Trace labels
$\lambda$ (of both our languages) have the following forms:
\[
\begin{array}{@{}l@{}l@{}}
\lambda & ~::=~ \tau \mid \checkmark \mid
\mathtt{ret}~?~\mi{Mem}, \mi{nalloc}
\mid
\mathtt{ret}~!~\mi{Mem}, \mi{nalloc}
\mid
\\
& 
\mathtt{call}(\mi{fid}) ~\OB{v}~?~\mi{Mem}, \mi{nalloc}
\mid
\mathtt{call}(\mi{fid}) ~\OB{v}~!~\mi{Mem}, \mi{nalloc}
\end{array}\]
\begin{list}{-}{\itemindent=0.5em \leftmargin=0em}
    \item A silent label $\tau$ abstracts over any execution step that
      is \emph{internal} to either the program \trg{p} or the context.
    \item A termination label $\checkmark$ indicates
    that a terminal execution state was reached.
    (Once a $\checkmark$ appears, it re-appears in all subsequent trace
    labels.)
    \item An input call label 
    $\mathtt{call}(\mi{fid}) ~\OB{v}~?~\mi{Mem}, \mi{nalloc}$
    indicates that at an execution state where the shared memory
    was $\mi{Mem}$, and the memory allocator state was $\mi{nalloc}$,
    the context called the program's 
    function
    $\mi{fid}$ with the list of values $\OB{v}$
    as arguments.
    \item An output call label
    $\mathtt{call}(\mi{fid}) ~\OB{v}~!~\mi{Mem}, \mi{nalloc}$
    is similar to an input call label but the call goes in the opposite direction:
    the program 
    called the context's function $\mi{fid}$.
    \item An input return label
    $\mathtt{ret}~?~\mi{Mem}, \mi{nalloc}$
    indicates that at an execution state where the shared memory
    was $\mi{Mem}$, and the allocator state was $\mi{nalloc}$,
    the context returned to the 
    program. 
    \item An output return label
    $\mathtt{ret}~!~\mi{Mem}, \mi{nalloc}$
    is similar except that the program 
    returned control to the context.
\end{list}

\begin{figure}[t]
\centering
\typerule{Return-to-program}{
    \trg{s \rightarrow s'} &
    \trg{s.M_{c}(s.pcc) = Return} 
    \\
    &
    \trg{s.pcc \nsubseteq \fun{dom}{\trg{p.M_{c}}}} & 
    \trg{s'.pcc \subseteq \fun{dom}{\trg{p.M_{c}}}}  \\
    \varsigma' = \fun{reachable\_addresses\_closure}
    {\varsigma, \trg{s'.Mem}}
    & \\
    \trg{Mem_{shr}} = \trg{s'.Mem}\vert_{\varsigma'}
}{(\trg{s}, \varsigma)~
\trsteptrg[p]{
        \neutralcol{\mathtt{ret}?~\trg{Mem_{shr}},~ 
            \trg{s.nalloc}}
    }~
(\trg{s'}, \varsigma')
}{return-to-program}
\caption{Trace semantics of
    \targetlanguage (Excerpt).
    The trace-step relation is indexed with a program \trg{p}.
}
\label{fig:trace-semantics}
\end{figure}

\Cref{fig:trace-semantics} shows the input return rule of
\trsteptrg[p]{\neutralcol{\lambda}}. The third and fourth
premises check that the program counter capability \trg{pcc} belongs
to $\trg{p}$'s code memory after the transition but not before,
implying that this is a border crossing from the context into the program of interest
$\trg{p}$. Similar checks exist in other rules (shown in the TR).



Next, given a reduction sequence of labeled steps, we drop all $\tau$
labels from it, and concatenate the non-$\tau$ labels into a
\emph{trace} $\alpha$, writing $(s, \varsigma)
\paronto{\neutralcol{\alpha}}_{p} (s', \varsigma')$ for the resulting steps.%
\footnote{These technicalities, all worked
  out in our TR, are mostly inspired by process
  calculi~\cite{milner1999communicating}.}
Then, we define the traces of a partial program $p$ as
in \Cref{def-tr}.  We note that all traces
are \emph{alternating} in ``$?$'' and ``$!$''.
\begin{fact}[Traces are alternating]
\label{assumption-traces-alternating}
$\alpha \in \mi{Tr}(p) \implies \alpha \in \mathtt{Alt} \checkmark^{*}$
~~where~~$\mathtt{Alt}~\defeq~(\overset{\bullet}{?}\vert \epsilon)~
(\overset{\bullet}{!}\overset{\bullet}{?})^{*}~
(\overset{\bullet}{!}\vert \epsilon)$
~~and~~$\overset{\bullet}{?}$ is the set of $?$-decorated
labels, and similarly for $\overset{\bullet}{!}$.
\end{fact}



\subsection{Proof of Lemma~\ref{no-trace-added} using \tricl}
\label{sec:explaining-tricl-intuition}

Lemma~\ref{no-trace-added} assumes a trace $\alpha$ for
$\compilersymbol{\src{p_s}}$ (in some target context, say,
$\trg{\ctx}$) and requires constructing a source emulating context
$\src{\ctx_{emu}}$ such that $\src{\ctx_{emu}[p_s]}$ also has
$\alpha$. For this, we back-translate $\alpha$ to a source context. We
illustrate the back-translation through an
example. \Cref{example-trace-prefix} shows one trace that is emitted
by the \emph{compiled} version of the module \src{Main} of
\Cref{src-lang-example}.

\begin{figure*}
\begin{alignat*}{2}
    \mathtt{call}
        \overbrace{(\mi{send\_rcv})}^{\mi{fid}}
        \overbrace{[(\delta, \sigma, \sigma + 512,0)]}^{\OB{v}}
        ~!~
&
        \underbrace{[\sigma \mapsto 0,\ldots,\sigma + 42 \mapsto 4242,\ldots,\sigma + 511 \mapsto 0]}_{\mi{Mem}}
        , \underbrace{-1}_{\mi{nalloc}}
\\
::~
    \mathtt{ret}&\
        ~?~
    \overbrace{[\sigma \mapsto 0,\ldots,\sigma + 42 \mapsto 0,\ldots,\sigma + 511 \mapsto 0]}
        ,~ \aoverbrace[L3U2R]{-1}
\end{alignat*}
\caption{Example trace of the compilation of the program in
  \Cref{src-lang-example}}
\label{example-trace-prefix}
\end{figure*}

The compilation of the first three commands generate $\tau$ steps,
which are dropped from traces.
The next two non-$\tau$ labels (shown in the example) are interesting:
\begin{enumerate}
\item the function call \src{send\_rcv(\&iobuffer)} on
  \Cref{line:callsend_rcv} is border crossing, so its compilation
  emits an output call label which contains the callee function id
  ($\mi{send\_rcv}$), the argument to the call (the
  $\delta$-capability representing the translation of the pointer
  \src{\&iobuffer}), the direction of the call ($!$ denoting output,
  i.e., program-to-context), a snapshot of the memory shared so far
  (namely, the contents of the array \src{iobuffer}), and the value
  $-1$ denoting the first heap address (the heap grows towards
  negative addresses in our semantics).
\item the target context (in which our compiled program executes)
  returns control to the compiled program, in this case,
    after \emph{zeroing out the contents of the shared memory}. This
    emits an input return label.
\end{enumerate}

\Cref{no-trace-added} requires showing that exactly the same trace
can be emitted by the source program in \emph{some} source
context. The proof of \Cref{no-trace-added}, therefore, requires us to
construct such a source context, which we call the
\emph{emulating} context. The emulating context depends on the
trace.\footnote{In principle, it could also depend on the target
context, but this is usually not required. We also don't use the
target context.}
\Cref{backtrans-example} shows an emulating context for our example trace.

\begin{lstlisting}[
caption= Example back-translation (simplified excerpt):
An \sourcelanguage~context emulating the trace
of \Cref{example-trace-prefix}.  , frame=tlrb,label=backtrans-example,
escapechar=|]
Module id: Networking
Import module: HelperBackTranslation

current_trace_idx; |\label{line:currenttraceidx_decl}|

send_rcv(iob_ptr) {
  Call readAndIncrementTraceIdx(&current_trace_idx); |\label{line:callreadandincr}|
  Call saveArgs_send_rcv_1(iob_ptr); |\label{line:callsaveargs}|
  Call saveSnapshot_0(); |\label{line:callsavesnapshot}|
  Call doAllocations_1(); |\label{line:calldoallocations}|
  Call mimicMemory_1(); |\label{line:callmimicmemory}|
  Return;  |\label{line:return}|
}
/*****************************************************/
Module id: HelperBackTranslation

current_trace_idx;
arg_store_0_send_rcv_0; |\label{line:argstorevariable}|
snapshot_0_|$\sigma$|;
|\ldots|
snapshot_0_|$\sigma + 511$|;

mimicMemory_1() {
  Assign *(arg_store_0_send_rcv_0[0]) 0;
  |\ldots|
  Assign *(arg_store_0_send_rcv_0[511]) 0;
  Return;
}
|\ldots|
\end{lstlisting}

This emulating source context consists of two modules,
\src{Networking}, which implements the API function \src{send\_rcv},
and \src{HelperBackTranslation}, which implements helper functions and
maintains metadata. We show just one example of such a helper
function, namely \src{mimicMemory\_1()}. \src{mimicMemory\_1()} is
called (on \Cref{line:callmimicmemory}) by \src{send\_rcv()}. It
zeroes out the IO buffer (to mimic the shared memory in the second
action of the given target trace). The IO buffer is accessed by
\src{mimicMemory\_1()} through the pointer stored in the global
variable \src{arg\_store\_0\_send\_rcv\_0}
(\Cref{line:argstorevariable}).  This pointer is stored (not shown) by
the function call \src{saveArgs\_send\_rcv\_1(iob\_ptr)} on
\Cref{line:callsaveargs}.

We briefly explain what each helper function does. First, the context
emulating a given trace defines a different set of helper functions
for every position on the trace.  The index of the corresponding trace
label appears in the identifier of a helper function (for
example, \src{mimicMemory\_1()}, and \src{saveSnapshot\_0()}). To
explain the helper functions, we follow the body
of \src{send\_rcv(iob\_ptr)} line by line.  In the beginning, the call
to \src{readAndIncrementTraceIdx} keeps track of the current position
in the trace. This knowledge of the current position in the trace is
not used in our toy example, but it would be used if the API function
(\src{send\_rcv} in this case) were called
\emph{at more than one position} in the given trace; at each
such position, we would use this knowledge to call the corresponding
helper functions (e.g., \src{mimicMemory\_3()} instead of
\src{mimicMemory\_1()}, which would copy to the shared memory the
values that appear in trace position 3 instead of~1).

Next, on \Cref{line:callsaveargs}, we store the pointer \src{iob\_ptr}
in a global variable by calling the \src{HelperBackTranslation} module
because we may need it to simulate a \emph{future} trace position, not
just the current call to \src{send\_rcv}. For the same reason, we save
(on \Cref{line:callsavesnapshot}) a snapshot of the whole shared
memory in global variables \src{snapshot\_0\_\sigma} to
\src{snapshot\_0\_\sigma+511}.

Next, on
\Cref{line:calldoallocations,line:callmimicmemory,line:return}, the
actual emulation of the trace action at trace position 1 is done. In
our example, \src{doAllocations\_1()} would do nothing because
\emph{nalloc} in \Cref{example-trace-prefix} does not
change. Importantly, \src{mimicMemory\_1()} writes all the values (the
zeros) to the shared memory before \src{send\_rcv} eventually returns,
thus mimicking the given target trace.

\paragraph*{\textbf{The difficult simulation}}
Returning back to the general picture of \Cref{no-trace-added}, after
we have constructed the emulating source context from the given trace
$\alpha$, we must \emph{prove} that the source program and this
context emulate the given target trace. As explained
in \Cref{sec:background}, for this, we would like to set up a
simulation between the target and source runs, but this simulation can
be very hard because there can be differences between
the \emph{internal} behaviors of the emulating context and the given
target context, e.g., in the specific order of updates to the shared
memory, and in the internal function calls. Indeed, the target context
likely does not use the kind of helper functions our emulating source
context does! Our simulation needs to accommodate this ``vertical
gap''.

To add to this difficulty, we do want to simulate through (the
$!$-decorated) execution steps of the program of interest (\src{p_s}
in \Cref{no-trace-added}).  And since our compiler compiles it, this
simulation would be very similar to what we would have to do anyhow
just to prove our compiler correct, not fully abstract. So, we would
like to ``reuse'' this part.

\paragraph*{\textbf{\tricl to the rescue}}
This is where our new \tricl simulation comes in. \tricl introduces a
third ``mediator'' run to the simulation, namely, that induced by the
\emph{compilation of the whole source program} \cmp{\src{\ctxs_{emu}[p_s]}}, containing both
\src{p_s} and our emulating context, say, \src{\ctxs_{emu}}.
As explained in \Cref{sec:background}, this simplifies the proof
because we can ``reuse'' whole-program compiler correctness as a black
box to immediately reduce the problem of showing that
the \emph{emulating} run emulates the given run to that of showing
that the \emph{mediating} run emulates the given run. The emulating
and mediating runs are in the same language, so this reduces the
``vertical gap''.

Finally, to show that the mediating run emulates the given run in the
target language, we rely on an \emph{alternating simulation} in the
target language that uses two different relations between the two runs
-- a \emph{strong} relation $\strongsim_{\compilersymbol{\src{p_s}}}$,
which holds while the compilation of the program of interest executes,
and a \emph{weak} relation $\weaksim_{\compilersymbol{\src{p_s}}}$,
which holds while the contexts \trg{\ctxt} (which produced $\alpha$)
and \cmp{\src{\ctxs_{emu}}} execute. The need for two relations will
become clear shortly.

\paragraph*{\textbf{\tricl formally}}
Formally, \tricl is a relation between three trace states (a source
emulating state \src{s_{emu}}, a target mediating state \trg{s_{med}}
and a target given state \trg{s_{given}} of the three runs explained
above) that agree on the memory shared between the context and the
program of interest. \tricl is indexed by a trace $\alpha$ that we are
trying to emulate and a position $i$ of that trace. The relation requires
that (1) $\src{s_{emu}}$ and $\trg{s_{med}}$ are related by the
whole-program compiler correctness relation
($\statecrossrelated_{\src{p_s}}$), (2) the source state satisfies an
\emph{emulation invariant}, which basically captures that the
construction of \src{\ctxs_{emu}} is indeed an emulation of
the \emph{input} steps of the trace $\alpha$, e.g., that the
functions \src{mimicMemory\_1()} and \src{doAllocations\_1()}
indeed emulate the input trace action $\alpha(1)$,
and (3) the two target states \trg{s_{med}} and \trg{s_{given}} are
related by the strong relation
$\strongsim_{\compilersymbol{\src{p_s}}}$ when execution is in
$\cmp{\src{p_s}}$ and by the weak relation
$\weaksim_{\compilersymbol{\src{p_s}}}$ when execution is in the
contexts.

\begin{definition}[Trace-Indexed Cross-Language 
	(\tricl) alternating simulation relation]\label{tricl-def}
	\begin{align*}
	&\fun{TrICL}{\src{s_{emu}}, \trg{s_{med}}, \trg{s_{given}},
		\varsigma}_{\alpha, i, \src{p_s}}
	~\defeq~\\
	&\qquad
    \src{s_{emu}} \statecrossrelated_{\src{p_s}} \trg{s_{med}}
	~\wedge~
    \fun{emulate\_invariants}{\src{s_{emu}}}_{\alpha, i,
		\src{p_s}}
	\\
    &\qquad \wedge~ 
    (\alpha(i) \in \overset{\bullet}{!} 
    \implies
	(\trg{s_{med}}, \varsigma)
	\strongsim_{\compilersymbol{\src{p_s}}} 
	(\trg{s_{given}}, \varsigma))
    \\
	&\qquad \wedge~ 
    (\alpha(i) \in \overset{\bullet}{?} \implies
	(\trg{s_{med}}, \varsigma) \weaksim_{\compilersymbol{\src{p_s}}} 
	(\trg{s_{given}}, \varsigma))
	\end{align*}
\end{definition}

Given this definition, we come to the key step of our proof, namely,
that \tricl is an invariant.
%
\begin{lemma}
	[\tricl step-wise alternating backward-simulation]
	\label{tricl-simulation-lemma}
	\begin{align*}
	&\alpha \in \mathtt{Alt}~\wedge~
    \fun{TrICL}{\src{s_{emu}}, \trg{s_{med}},
		\trg{s_{given}}, \varsigma}_{\alpha, i, \src{p_s}}~\wedge \\
    &(\trg{s_{given}}, \varsigma) 
	~\trg{\paronto{\neutralcol{\alpha(i)}}_{\compilersymbol{\src{p_s}}}}~
	(\trg{s_{given}'}, \varsigma')
    \\
	\implies~
	&\exists \src{s_{emu}'}, \trg{s_{med}'}.~
    (\src{s_{emu}}, \varsigma)
	~\src{\paronto{\neutralcol{\alpha(i)}}_{p_s}}~
	(\src{s_{emu}'}, \varsigma')~\wedge\\
    &(\trg{s_{med}}, \varsigma) 
	~\trg{\paronto{\neutralcol{\alpha(i)}}_{\compilersymbol{\src{p_s}}}}~
	(\trg{s_{med}'}, \varsigma')~\wedge
    \\
	&
    \fun{TrICL}{\src{s_{emu}'}, \trg{s_{med}'},
		\trg{s_{given}'}, \varsigma'}_{\alpha, i+1, \src{p_s}}\\
	\end{align*}
\end{lemma}

Before sketching the proof of this key lemma,
we explain how the proof reuses the backward-
and forward- simulation lemmas that are anyhow needed for compiler
correctness:
\begin{itemize}
\item
When control is in the contexts (case
$\alpha(i) \in \overset{\bullet}{?}$ of \Cref{tricl-def}), we know by
the \emph{emulation invariant} that the source context
$\src{\ctxs_{emu}}$ emulates the next action of the trace $\alpha$
(the emulation invariant holds of the construction shown in the
example earlier), and we use \emph{forward simulation} to argue that
the mediating context, which is just the compilation of this source
context, does the same.
\item  Dually, when control is in the program of interest 
(case $\alpha(i) \in \overset{\bullet}{!}$ of \Cref{tricl-def}), we
know by the precondition of \Cref{tricl-simulation-lemma} that
$\cmp{\src{p_s}}$ produces the next action of the
\emph{given} trace $\alpha$, hence (by strong similarity that we explain below)
$\cmp{\src{p_s}}$ also produces the same action in the
\emph{mediator} trace. But now from knowing that an action of the mediator trace
was produced,
we use \emph{backward simulation}
to reason that the source program $\src{p_s}$ does the same action.
\end{itemize}
 This ``reuse'' of the (forward- and backward-) simulations that are
anyhow needed for compiler correctness is the simplification
that \tricl affords.  The proof
of the two interesting cases of
\Cref{tricl-simulation-lemma} is given in \Cref{appendix:assms-tricl}.%
\footnote{The proof relies on forward-
and backward-simulation, and also on key properties of the strong and
weak similarities and of the emulate invariants. All of these
properties
(\Cref{assm-fwdsim,assm-bckwdsim,assm-lock-step-simulation,assm-weakening,assm-option-simulation,assm-strengthening,assm-emul-adequacy,assm-emul-invar})
are given in \Cref{appendix:assms-tricl}, and are proved in the
TR.}

\begin{figure}
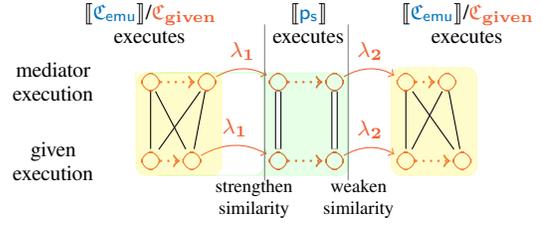

    \centering
    \tikzpic{
    \node[trgstate](sc1){};
    \node[trgstate, right = .5 of sc1](sc2){};
    \node[trgstate, right = .7 of sc2](sc3){};

    \node[trgstate, below = .8 of sc1](tc1){};
    \node[trgstate, right = .3 of tc1] (tc2){};
    \node[trgstate,] at(sc3|-tc1) (tc3){};


    \node[trgstate, right = .5 of tc3](tc6){};
    \node[trgstate, right = .7 of tc6](tc7){};
    \node[trgstate, right = .5 of tc7](tc8){};

    \node[trgstate,] at(tc6|-sc1) (sc6){};
    \node[trgstate,] at(tc7|-sc1) (sc7){};
    \node[trgstate, right = .3 of sc7] (sc8){};

    \draw[trgredstar] (sc1) to (sc2);
    \draw[trgred] (sc2) to node[above,font=\footnotesize](as1){\trg{\lambda_1}} (sc3);
    \draw[trgredstar] (sc3) to (sc6);
    \draw[trgred] (sc6) to node[above,font=\footnotesize](as2){\trg{\lambda_2}} (sc7);
    \draw[trgredstar] (sc7) to (sc8);

    \draw[trgredstar] (tc1) to (tc2);
    \draw[trgred] (tc2) to node[above,font=\footnotesize](at1){\trg{\lambda_1}} (tc3);
    \draw[trgredstar] (tc3) to (tc6);
    \draw[trgred] (tc6) to node[above,font=\footnotesize](at2){\trg{\lambda_2}} (tc7);      
    \draw[trgredstar] (tc7) to (tc8);


    \draw[rounded corners, draw=yellow, fill=yellow, opacity = .2 ] (sc1.north) -| ([xshift=.2em]sc2.east) |- ([yshift=-.2em]tc2.south) -| ([xshift=-.2em]sc1.west) |- (sc1.north);
    \draw[rounded corners, draw=yellow, fill=yellow, opacity = .2 ] ([yshift=.2em]sc7.north) -| ([xshift=.2em]tc8.east) |- ([yshift=-.2em]tc8.south) -| ([xshift=-.2em]sc7.west) |- ([yshift=.2em]sc7.north);

    \draw[rounded corners, draw=green, fill=green, opacity = .1 ] (sc1.north) -| ([xshift=.2em]sc6.east) |- ([yshift=-.2em]tc1.south) -| ([xshift=-.2em]sc3.west) |- (sc1.north);


    \draw[-, draw=gray] ([yshift=1em]as1.east) -- ([yshift=-5em]as1.east);
    \draw[-, draw=gray] ([yshift=1em]as2.west) -- ([yshift=-5em]as2.west);

    \node at([yshift=-5.6em,xshift=-.4em]as1.east) [align=center,font=\scriptsize](){strengthen \\ similarity} ;
    \node at([yshift=-5.6em,xshift=.4em]as2.west) [align=center,font=\scriptsize](){weaken \\ similarity} ;

    \node[font = \footnotesize,above = .4 of sc3.east, align = center, xshift=.8em](y1){\cmp{\src{p_s}} \\ executes};
    \node[font = \footnotesize, left = .5 of y1, align = center](yl){\cmp{\src{\ctxs_{emu}}}/\trg{\ctxt_{given}} \\ executes};
    \node[font = \footnotesize,right = .5 of y1, align = center](yr){\cmp{\src{\ctxs_{emu}}}/\trg{\ctxt_{given}} \\ executes};

    \node[font = \footnotesize, left = .5 of sc1, align = center](yr){mediator \\ execution};
    \node[font = \footnotesize, left = .5 of tc1, align = center](yr){given \\ execution};

    \draw[staterelweak,] (sc1) to (tc1);
    \draw[staterelweak,] (sc1) to (tc2);
    \draw[staterelweak,] (sc2) to (tc2);
    \draw[staterelweak,] (sc2) to (tc1);

    \draw[staterelstrong] (sc3) to (tc3);
    \draw[staterelstrong] (sc6) to (tc6);
    
    \draw[staterelweak,] (sc7) to (tc7);
    \draw[staterelweak,] (sc7) to (tc8);
    \draw[staterelweak,] (sc8) to (tc8);
    \draw[staterelweak,] (sc8) to (tc7);

}

    \caption{Alternating strong and weak relations between the
    	`mediator' execution (top) and the `given' execution
    	(bottom).}  \label{fig:strong-weak-general}
\end{figure}

\paragraph*{\textbf{The strong and weak relations}}
We now explain the strong relation
$\strongsim_{\compilersymbol{\src{p_s}}}$ and the weak
relation~$\weaksim_{\compilersymbol{\src{p_s}}}$ between the target
trace states of the mediating run (of \cmp{\src{\ctxs_{emu}[p_s]}}) and the
given run (of \trg{\ctxt[\cmp{\src{p_s}}]}).
Following \Cref{tricl-def}, the strong relation holds between trace
states while control is in the compilation of the program of interest
\cmp{\src{p_s}} in the two runs. Since this program (part) is exactly the
same in both runs,
 the strong relation is a lock-step simulation (\Cref{assm-lock-step-simulation}), which maintains
similarity of the call stacks, the private memory of the program of
interest and the shared memory (memory accessible to both the context
and the program of interest).

In contrast, the weak relation holds between trace states while
control is in the contexts \cmp{\src{\ctxs_{emu}}} and
\trg{\ctxt}. Since these two contexts can be very different, this
relation is not lock-step. It allows one step on either side to be
simulated by zero or more steps on the other side (\Cref{assm-option-simulation}) and allows a
different sequence of internal function calls (and, hence, stack
states). In terms of memory, it only enforces that the \emph{private
  memory of the program of interest} remain in sync, but the shared
memory and the private memory of the contexts may diverge arbitrarily.

In the proof of \Cref{tricl-simulation-lemma}, these relations must be
replaced by each other at border crossings.
At a border crossing where control transfers from the program of
interest to the contexts (case $\alpha(i) \in \overset{\bullet}{!}$),
we must ``weaken'' the strong relation to the weak one. This is quite
straightforward since the strong relation directly implies the weak
relation.

In the other direction, at a border crossing where control
transfers from the contexts to the program of interest 
(case $\alpha(i) \in \overset{\bullet}{?}$), we must
``strengthen'' the weak relation to the strong relation. For this, we
must prove that the shared memory (recorded on $\alpha(i)$) is exactly the same in the two
target runs at that point. This is not obvious and is proved as
follows. First, we know from the \tricl invariants that the third
trace (the emulating source trace) is mimicking the step
$\alpha(i)$.
We then use this conclusion to show that the trace label $\alpha(i)$ at the
border crossing in question must be the same in the given and the
mediator runs (the two target runs). Here is how:
By applying compiler-correctness forward simulation to
the emulating run so far, we conclude that
the mediating trace must also take a step with the same label
$\alpha(i)$. Since the shared memory is recorded in the label
$\alpha(i)$, it immediately follows that the shared memory is exactly
the same in the three runs, and in particular in the two target
runs, which completes the strengthening proof.

\Cref{fig:strong-weak-general} depicts the alternating 
nature of the strong and weak relations, and the strengthening and
weakening at border crossings. The two traces are depicted as two
horizontal sequences of states/transitions. The black lines that
connect states from opposite traces show the nature of the simulation
condition: option simulation (\Cref{assm-option-simulation}) is
possible for weak similarity (the single black line), while for strong
similarity (the double black line), only lock-step simulation
(\Cref{assm-lock-step-simulation}) is possible.

\section{Implementation in a compiler from C to CHERI}
\label{sec:implementation}



We have implemented the key ideas of \Cref{sec:compiler} in a compiler
from C to \cheri. \cheri comes with a Clang/LLVM
compiler~\cite{chericompartment} that already implements \pac, but no
isolation for modules, which \sourcelanguage has. Our compiler
enforces this isolation via a C-to-C source pre-processing
transform. The compiler relies on \texttt{libcheri}, CheriBSD's
library for building, invoking and loading \emph{sandboxes} --
isolated units of computation with their own code and memory, which is
private until explicitly shared. \texttt{libcheri} relies on \cheri's
underlying support for object capabilities to provide
sandboxes~\cite{cheriprogrammers}. The important thing from our
perspective is that \texttt{libcheri} requires the programmer to
manually group functions into classes (the equivalent of modules), and
to annotate functions to make them use object capabilities instead of
the standard calling conventions (attribute \lstinline|cheri_ccall|)
and to specifically annotate functions that are exported from a class
(attribute \mbox{\lstinline|cheri_ccallee|}). Additionally, all
initialization functions must be added to classes that call external
functions.

Our source-to-source transform automates all this: It maps C modules
(compilation units) to \texttt{libcheri}'s sandboxes to isolate them
from each other and automatically inserts all the required
annotations. Examples of programs output by our transform are shown in
the TR. To initialize sandboxes, we had to make some significant
changes to \texttt{libcheri} as well. Briefly, \texttt{libcheri}
requires a second phase of runtime linking to resolve cross-sandbox
references. We implement this through a new recursive initialization
function, called once before \lstinline{main}, which loads and
initializes all sandboxes that \mbox{\lstinline{main}} depends on
transitively, creates relevant object capabilities, and links the
modules.

\paragraph*{Differences from the formalization}
Our idealized source language \sourcelanguage provides accessors for
bounds information about pointers whereas native C, that our
implementation supports, does not. \cheri's C interface, however,
provides pointers with such accessors. (These pointers wrap \cheri
capabilities.) We rely on the programmer to use these \cheri pointers
in place of native C pointers consistently. We have also not yet
implemented the (straightforward) link-time check on target contexts
to prevent them from accessing \trg{pcc} directly, something that our
ideal target language \targetlanguage enforces.

\section{Experimental evaluation}
\label{sec:experiments}

We evaluate the overheads of our proof-of-concept isolation scheme
using four large open-source C libraries that we chose carefully for
heterogeneity and compiled with our compiler:
\texttt{zlib}~\cite{refzlib}, \texttt{LibYAML}~\cite{refyaml},
\texttt{GNU-barcode}~\cite{refbarcode}, and
\texttt{libpng}~\cite{refpng}.



\begin{table}
\begin{center}
    \begin{tabular}{| l | l | l | l | l |}
    \hline
    Software & \texttt{zlib} & \texttt{LibYAML} & \texttt{GNU-barcode} & \texttt{libpng} \\ \hline\hline
    & \multicolumn{4}{|c|}{Porting overhead summary}\\ \hline
    Lines of code & 11255 & 12762 & 4657 & 33029 \\ \hline
    Altered lines & 130 & 114 & 164 & 51 \\ \hline
    Percentage & 1.15\% & 0.89\% & 3.5\% & 0.15\%  \\ \hline\hline
    & \multicolumn{4}{|c|}{No.\ of occurrences of each incompatible pattern}\\
    \hline
    Passed local ref & 2 & 15 & 0 & 13 \\ \hline
    Extern global var & 3 & 0 & 0 & 0 \\ \hline
    Pointer to ext fun & 2 & 1 & 26 & 2 \\ \hline
    Other & 0 & 1 & 1 & 4 \\ \hline
    Total & 7 & 17 & 27 & 19 \\ \hline
    \end{tabular}
\end{center}
\caption{Porting overheads}
\label{table:modifications}
\end{table}



\paragraph*{Code changes}
We had to make some manual code changes to address incompatibilities
between our scheme and the existing toolchain. These incompatibilities
are: (a)~Moving local variables whose addresses are shared with other
modules to the heap; this is a fundamental limitation of \cheri, not
specific to our scheme. (b)~Explicitly sharing pointers to exported
global variables of a module; this can be automated with further work
on the \cheri linker. (c)~In the existing toolchain, a function
pointer always compiles to an execute capability. Our isolating scheme
requires object capabilities for cross-module calls. To compensate for
this incompatibility, we had to make manual changes to cross-module
calls using function pointers.
\Cref{table:modifications} summarizes the magnitude of our changes on
the various benchmarks. Overall, the changes are quite limited.



\begin{figure*}[t]
  \centering
  \begin{subfigure}[b]{.26\textwidth}
    \centering
    \caption{\texttt{zlib}}
    \label{subfig:zlib}
    \begin{tikzpicture}[scale=0.55]
      \begin{semilogxaxis}[
        xlabel={Payload size (MB)},
        ylabel={Time (s)},
        legend entries={secure, vanilla},
        legend pos=north west]
        \addplot table [x=size, y=secc, col sep=comma] {./resources/zlib-times.csv};
        addlegendimage{}
        \addplot table [x=size, y=mabi, col sep=comma] {./resources/zlib-times.csv};
        addlegendimage{}
      \end{semilogxaxis}
    \end{tikzpicture}
  \end{subfigure}
  \hfill
  \begin{subfigure}[b]{.26\textwidth}
    \centering
    \caption{\texttt{LibYAML}}
    \label{subfig:yaml}
    \begin{tikzpicture}[scale=0.55]
      \begin{semilogxaxis}[
        xlabel={Payload size (MB) },
        ylabel={Time (s)},
        legend entries={secure, vanilla},
        legend pos=north west]
        \addplot table [x=size, y=secc, col sep=comma] {./resources/yaml-times-c.csv};
        addlegendimage{}
        \addplot table [x=size, y=mabi, col sep=comma] {./resources/yaml-times-c.csv};
        addlegendimage{}
      \end{semilogxaxis}
    \end{tikzpicture}
  \end{subfigure}
  \hfill
  \begin{subfigure}[b]{.26\textwidth}
    \centering
    \caption{\texttt{libpng}}
    \label{subfig:libpng}
    \begin{tikzpicture}[scale=0.55]
      \begin{semilogxaxis}[
        xlabel={Image height (rows) },
        ylabel={Time (s)},
        legend entries={secure, vanilla},
        legend pos=north west]
        \addplot table [x=size, y=secc, col sep=comma] {./resources/png-times-c.csv};
        addlegendimage{}
        \addplot table [x=size, y=mabi, col sep=comma] {./resources/png-times-c.csv};
        addlegendimage{}
      \end{semilogxaxis}
    \end{tikzpicture}
  \end{subfigure}
  \caption{Performance benchmark results. Note that x-axes are log-scale.}
  \label{fig:benchmarks}
\end{figure*}
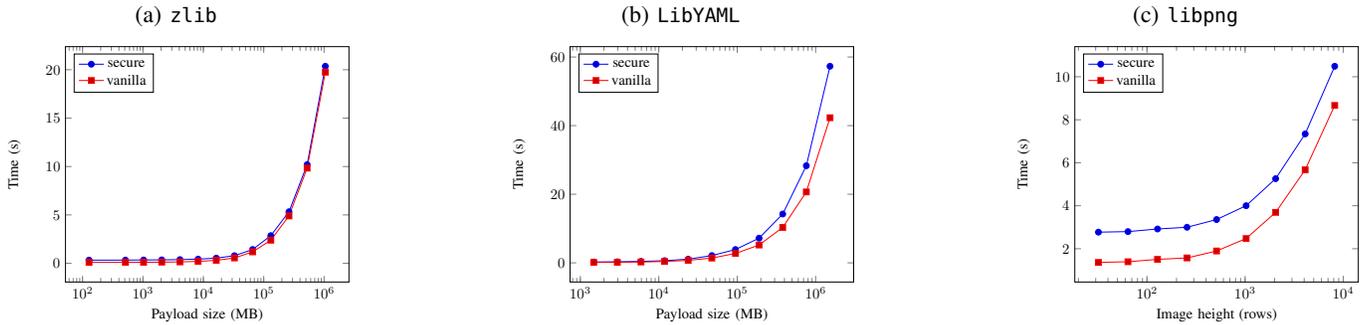



\paragraph*{Performance overheads}
The most significant performance overhead in our scheme comes from
cross-module function calls, where we use the \cheri
\mbox{\lstinline{ccall}} instruction, which is used to implement
object capabilities. This instruction is expensive as it performs a
number of operations like saving and restoring callee-save registers,
clearing unused argument registers and making certain checks on
arguments.\footnote{Some of this overhead is \emph{not} fundamental
  and the \cheri design has improved on \mbox{\lstinline{ccall}} over
  time~\cite{micro16}. The overhead reported here is based on
  \mbox{\lstinline{ccall}} as it exists in \cheri ISA version 5, and
  should be considered conservative.}
The number of cross-module calls is, of course, application
dependent. To isolate this overhead, we compare the execution times of
calls to library functions compiled with our scheme to those compiled
with the vanilla \cheri compiler, which offers no module isolation
and, hence, no security. We perform our experiments on a single-core
\cheri VM implementing \cheri ISA version 5 \cite{cheriisatechv5}
running our modified version of CheriBSD. This VM was hosted in
another VirtualBox~\cite{10.5555/2613694} VM with FreeBSD 9.1, which,
in turn, runs on an arch-Linux host. The physical machine has a 4 core
Intel Core i7-6700 CPU with 16 GB of RAM, of which 512 MB are
allocated to the \cheri VM.

\Cref{fig:benchmarks} summarizes our results. The y-axes are execution
times. The x-axes are log-scale. The lines ``secure'' and ``vanilla''
correspond to our scheme and vanilla \cheri, respectively. All times
are averages of 10 runs. Standard deviations were all below 0.3s with
the exception of the two longest \texttt{LibYAML} experiments, where
they were 1.5s in each case. A small, constant delay of about 0.2s
associated with the sandbox loading routines is included in both the
baseline and the secure versions. We do not show the benchmark
\texttt{GNU-barcode} as it takes very small inputs only.

Our overhead relative to vanilla \cheri is negligible for
\texttt{zlib} as this benchmark has very few cross-module calls
(43--441 as the payload size varies). In \texttt{LibYAML}, the number
of cross-module calls grows linearly with input size, so the
\emph{ratio} of the two lines is nearly constant (our relative
overhead is consistently between 35 and 40\%). In \texttt{libpng}, the
number of cross-module calls is constant but large (approx.\ 125,000),
so the \emph{difference} between the two lines is nearly constant and
easily perceivable. Overall, these observations are in line with our
expectation that the performance overhead of our isolation scheme is
dominated by the number of cross-module calls and that the overhead of
each such call is constant.

The absolute numbers reported here should not be extrapolated to other
hardware as they were obtained on a \cheri simulator that is not cycle
accurate. Nonetheless, we expect similar \emph{trends} to hold, i.e,
the overhead should be roughly linear in the number of cross-module
calls on all similar hardware. Another noteworthy point is that a
compiler may choose a different isolation boundary, e.g, it could
isolate individual class objects to prevent access to their private
fields from outside. In such a case, the primary overhead would shift
to crossings of that isolation boundary. In the said example, the
overhead would become proportional to the number of cross-object
method invocations.



\section{Related work}\label{sec:related-work}
\paragraph*{Verified compilation to capability machines}
The work closest to ours
is \emph{StkTokens}~\cite{Skorstengaard:2019:SEW:3302515.3290332} as
it also verifies a compiler transformation that targets capability
machines. However, \emph{StkTokens} and our work are complementary:
While we model and prove secure the \pac
transformation, \emph{StkTokens} models and verifies a transformation
that implements a calling convention. To do this, \emph{StkTokens}
assumes a hardware extension called \emph{linear capabilities}, for
which experimental support exists in \cheri. We believe
that \emph{StkTokens} can be combined with our work to eliminate our
trust in the control stack. 
Van Strydonck et al~\cite{DBLP:journals/pacmpl/StrydonckPD19}
also use linear capabilities.
An important difference is that Van Strydonck et al's compiler relies on static contracts defining which memory is shared across trust boundaries, while ours
 considers components which share memory dynamically during execution.

\paragraph*{Full abstraction for compiler security}
The idea of using full abstraction to formalize compiler security was
introduced by Abadi~\cite{abadiprotectiontranslations}. Compiler full
abstraction (or security in general) is typically proved either with a trace-directed
back-translation~\cite{catalin,Abate:2018:GCG:3243734.3243745,patrignani2015secure,Patrignani:2019:FAS:3303862.3280984} or a syntax-directed
back-translation~\cite{New,DBLP:journals/lmcs/DevriesePPK17,Skorstengaard:2019:SEW:3302515.3290332,DBLP:journals/pacmpl/StrydonckPD19,AhmedFa,ahmedCPS}. A
syntax-directed back-translation defines a reverse compiler that
translates the assumed distinguishing target context (a piece of
syntax) to a source context. While this approach has been used
extensively in prior work, it can be difficult to use in situations
where some target-level constructs have no obvious source
counterparts. This is not the case for our languages, but we follow
the trace-directed approach as we still find it technically
simpler. We contribute to the trace-directed technique by
introducing \tricl, which allows reuse of whole-program compiler
correctness to simplify the proof of the ``security direction'' of
full abstraction
(\Cref{theorem-full-abstraction}(\ref{condition-preserves})).  We are
not aware of similar reuse of whole-program compiler correctness in
the syntax-directed method.

\paragraph*{Other criteria for secure compilation}
A recent line of work proposes alternate criteria for compiler
security, based on preservation of classes of properties and
hyperproperties in the presence of adversarial
contexts~\cite{abate2019journey,rsc,Abate:2018:GCG:3243734.3243745,schp,DBLP:conf/esop/AbateBCD0HPTT20}. Many
of these criteria are incomparable to full abstraction, while some are
stronger. Work on proof techniques for proving these criteria is still
in early stages, but back-translation (both trace-directed and
syntax-directed) features prominently. In particular, Abate \emph{et
al.}~\cite{Abate:2018:GCG:3243734.3243745} describe a compiler and
prove a strong robust safety property (not \fa) for it using
trace-directed back-translation. Their method also reuses
whole-program compiler correctness to reduce a big part of cross-language reasoning
to only target-level reasoning. However, their setting does not
support sharing of memory across modules, which substantially
simplifies their proof by eliminating the need for the strong and weak
relations (\Cref{sec:explaining-tricl-intuition}).

Another line of
work~\cite{10.1145/3371075,barthe2018secure,sison_et_al:LIPIcs:2019:11082,besson2018securing,watt2019ct}
verifies that the compiler does not undo countermeasures that the
programmer of cryptographic libraries implements in order to ensure
protection against timing attacks or other secret-revealing attacks.

\paragraph*{Fully-abstract trace semantics}
Our trace labels are inspired by Laird's work for a functional
language with general
references~\cite{9249fdc52dd3414e83b2e3f9a89cb117}. Laird relies on
a \textit{bipartite} LTS in which nodes are partitioned between
program configurations and environment (context) configurations.  We
do not use this segregation explicitly, but our checks on \trg{pcc} in
the trace semantics have the same effect.

\section{Limitations and Future Work}\label{sec:limitations}
Our work shows formally that \pac compilers can provide strong
guarantees for partial programs. While we believe that this is a
significant step forward in the understanding of the security
properties of \pac compilers, we still make some simplifications and
assumptions that would be interesting to remove in future work. First,
our memory allocation model does not support de-allocation. This
simplification allows us to represent the state of the memory
allocator as just the next-free-address, and this is essential in
keeping our model manageable. To the best of our knowledge, nobody has
yet developed fully abstract trace semantics for languages with a
realistic model of deallocation. Work in this direction would be
interesting. Second, we do not yet model side channels. As such, our
compiler is not guaranteed to preserve resistance against side-channel
leaks~\cite{d2015correctness}. There is recent related work that
specifically investigates how to secure compilers such that they
preserve side-channel
resistance~\cite{10.1145/3371075,besson2018securing,watt2019ct,barthe2018secure,sison_et_al:LIPIcs:2019:11082}. Combining \pac
with these ideas would also be interesting.

\paragraph*{Acknowledgments}
We thank several anonymous reviewers for their useful feedback. This
research was partially funded by the Research Fund KU Leuven, by the
Research Foundation - Flanders (FWO), by a gift from Intel
Corporation, and by the German Federal Ministry of Education and
Research (BMBF) through funding for the CISPA-Stanford Center for
Cybersecurity (FKZ: 13N1S0762).

\bibliographystyle{IEEEtranN}
\bibliography{main}

\appendix
\section{Appendix}




\subsection{Some details of the proof of \Cref{tricl-simulation-lemma}}
\label{appendix:assms-tricl}

We show some excerpted definitions of key constituents of our \tricl
relation and the proof of \Cref{tricl-simulation-lemma}, the step-wise
backward simulation condition of \tricl, which is an interesting proof
because it puts together the whole-program compiler correctness
lemmas.
We group the (discharged) assumptions used in the proof
of \Cref{tricl-simulation-lemma}
(namely, \Cref{assm-fwdsim,assm-bckwdsim,assm-lock-step-simulation,assm-weakening,assm-option-simulation,assm-strengthening,assm-emul-adequacy,assm-emul-invar}
below) by the components of \tricl they refer to. We recall
that \tricl is defined in terms of four main relations/invariants:
\begin{enumerate}
\item the vanilla (whole-program) 
\textbf{compiler-correctness relation}
($\statecrossrelated_\src{p}$)
between the source state and the mediator
 state
satisfying lifted forward- and backward-simulations
(\textbf{\Cref{assm-fwdsim,assm-bckwdsim}})

\item a \textbf{strong-similarity relation} 
($\strongsim_{\trg{p}}$)
between the mediator
state and the given state, a relation that satisfies 
lock-step simulation (\textbf{\Cref{assm-lock-step-simulation}})

\item a \textbf{weak-similarity relation} 
($\weaksim_{\trg{p}}$) also between
the mediator
state and the given state, a relation that satisfies 
option simulation (\textbf{\Cref{assm-option-simulation}}),
and -- together with the strong similarity -- satisfying
both weakening (\textbf{\Cref{assm-weakening}}) and
strengthening (\textbf{\Cref{assm-strengthening}})

\item \textbf{emulation invariants} about the source
state satisfying both adequacy (\textbf{\Cref{assm-emul-adequacy}})
and preservation by trace steps (\textbf{\Cref{assm-emul-invar}})
\end{enumerate}

\begin{definition}[Weak-similarity relation (excerpt)]
	\label{weak-rel-excerpt}
~\\$(\trg{s_1}, \varsigma) \weaksim_{\trg{p}}
(\trg{s_2}, \varsigma)~\defeq\ 
\trg{s_1.stk} \weaksim_{\trg{p}} \trg{s_2.stk}\ \wedge\ 
\trg{s_1.mem}\ \weaksim_{\trg{p}, \varsigma}\ \trg{s_2.mem}
$\\
where weak stack similarity and weak memory similarity
are defined as follows:\\
$\trg{s_1.stk} \weaksim_{\trg{p}} \trg{s_2.stk}$ is defined
to be the existence of a map between the indices of the
entries of \trg{s_1.stk} and the indices of the entries
of \trg{s_2.stk}, such that this map is ``successor preserving'',
monotone, and every pair of indices that is in the map
corresponds to two equal stack entries (one from \trg{s_1.stk}
and the second from \trg{s_2.stk}).
(See section 6.3 of the technical report for formal definitions.)\\
$\trg{s_1.mem} \weaksim_{\trg{p}, \varsigma} \trg{s_2.mem}$ 
is defined to be equality of the contents of memories 
\trg{s_1.mem} and \trg{s_2.mem} at
all the currently-private addresses of \trg{p}, where a 
currently-private
address of \trg{p} is any address that is so far not shared 
(i.e., not
in the set $\varsigma$) and that is reachable from \trg{p}'s 
statically-allocated memory.
\end{definition}

\begin{definition}[Strong-similarity relation (excerpt)]
	\label{strong-rel-excerpt}
~\\$(\trg{s_1}, \varsigma) \strongsim_{\trg{p}}
(\trg{s_2}, \varsigma)~\defeq\ 
\trg{s_1.stk} \strongsim_{\trg{p}} \trg{s_2.stk}\ \wedge\ 
\trg{s_1.mem}\ \strongsim_{\trg{p}}\ \trg{s_2.mem}
$
where strong stack similarity and strong memory similarity
are defined as follows:\\
$\trg{s_1.stk} \strongsim_{\trg{p}} \trg{s_2.stk}$ is defined
to be the same as weak stack similarity, i.e.,
the existence of a map with the same properties
mentioned in \Cref{weak-rel-excerpt}, but with the extra
condition that the top-most stack indices must be in the map.
(This extra condition ensures that all function calls
and returns happen in sync, hence, satisfying
the lock-step simulation condition of
 \Cref{assm-lock-step-simulation}.)\\
$\trg{s_1.mem} \strongsim_{\trg{p}} \trg{s_2.mem}$ 
is defined to be equality of the contents of memories 
\trg{s_1.mem} and \trg{s_2.mem} at
all the addresses reachable from \trg{p}'s 
statically-allocated memory.
\end{definition}

\begin{definition}[Emulation invariants (simplified excerpt)]
	\label{emulate-invariants-excerpt}
The emulation invariants for a state \src{s}
and the prefix of a finite trace $\alpha$
up to position $i$ consist of invariants on the
shape of the code that is currently executing (i.e., the code
starting at 
address \src{s.pc}), in addition to
invariants on memory \src{s.mem}
that ensure the metadata (that the back-translated
context keeps) is compatible with
the information from the trace prefix of $\alpha$ up to position
$i$.
The invariants on memory can be found in the technical report
in definitions 116, 117, 118, and 126.
The emulation
invariants on the shape of the code are more interesting:
~\\$\fun{emulate\_invariants}{\src{s}}_{\alpha, i, \src{p}}\ \defeq\ \\
\alpha(i) \in \overset{\bullet}{?} \implies 
\exists j.~j \leq i\ \wedge\ 
\mathtt{upcoming\_commands}(
\src{s},\\ 
\fun{emulate\_responses\_for\_suffix}{\alpha, j, \src{s.pc}})
$\\
where:\\
$\mathtt{upcoming\_commands}(\src{s}, c)$ ensures that 
starting at address \src{s.pc} of the code memory of \src{s},
the commands $c$ are allocated, and\\
$\mathtt{emulate\_responses\_for\_suffix}(\alpha, j, \ldots)$
is defined recursively on $j$ to be a nested switch statement:
$\mathtt{switch}(\src{current\_trace\_idx})\ \{$\\
$\mathtt{case}~j: \\
~~~~~~~~~~~~\mathtt{emulate\_ith\_action}(\alpha, j, 
\ldots); \\
~~~~~~~~~~~~\mathtt{emulate\_responses\_for\_suffix}(\alpha, j + 2, \ldots)$\\
$\mathtt{case}~j+2: \\
~~~~~~~~~~~~\mathtt{emulate\_ith\_action}(\alpha, j+2, 
\ldots); \\
~~~~~~~~~~~~\mathtt{emulate\_responses\_for\_suffix}(\alpha, j + 4, \ldots)$\\
$\ldots$\\
$\mathtt{case}~\lvert\alpha\rvert - 1: \\
~~~~~~~~~~~~\mathtt{emulate\_ith\_action}(\alpha, \lvert\alpha\rvert - 1, 
\ldots);$\\
$\}$\\
(See definitions 120 and 121 
in the technical report for
the formal definitions.)\\
(Notice that
the maximum depth of the nesting of the
switch statement
is $(\lvert \alpha \rvert - j) / 2$, 
i.e., half the length of the suffix starting
at position $j$---the half originates
from the fact that we are only emulating every other
trace label.)\\
(Notice also that, interestingly, most of the code generated
by $\mathtt{emulate\_responses\_for\_suffix}$ is actually
unreachable; for a finite trace $\alpha$,
the size of the code generated is $O(\lvert\alpha\rvert^2)$,
of which only $O(\lvert\alpha\rvert)$
is reachable. However, this extra code simplifies
some proofs.)\\
(Observe that 
 $\mathtt{emulate\_ith\_action}$ 
 generates the code in the back-translation
 example of \Cref{backtrans-example}---\cref{line:callreadandincr,line:callsaveargs,line:callsavesnapshot,line:calldoallocations,line:callmimicmemory,line:return})
\end{definition}


\begin{lemma}[Compiler
     forward-simulation lifted to compressed trace steps]
    \label{assm-fwdsim}
$ 
\src{s_s} \statecrossrelated_\src{p} \trg{s_t} \wedge 
(\src{s_s}, \varsigma) ~\src{\paronto{\neutralcol{\alpha}}_{p}}~
(\src{s_s'}, \varsigma')
~\implies~ \exists \trg{s_t'}.~(\trg{s_t}, \varsigma)
~\trg{ \paronto{\neutralcol{\alpha}}_\neutralcol{\compilersymbol{\src{p}}}}~
(\trg{s_t'}, \varsigma')
 \wedge
\src{s_s'} \statecrossrelated_\src{p} \trg{s_t'}
$
\end{lemma}

\begin{lemma}[Compiler
    backward-simulation lifted to compressed trace steps]
    \label{assm-bckwdsim}
$ 
\src{s_s} \statecrossrelated_\src{p} \trg{s_t} \wedge 
(\trg{s_t}, \varsigma)
~\trg{ \paronto{\neutralcol{\alpha}}_\neutralcol{\compilersymbol{\src{p}}}}~
(\trg{s_t'}, \varsigma')
~\implies~ \exists \src{s_s'}.~
(\src{s_s}, \varsigma) ~\src{\paronto{\neutralcol{\alpha}}_{p}}~
(\src{s_s'}, \varsigma')
\wedge
\src{s_s'} \statecrossrelated_\src{p} \trg{s_t'}
$
\end{lemma}

\begin{lemma}[Lock-step simulation of strong similarity]
    \label{assm-lock-step-simulation}
~\\$ 
(\trg{s_1}, \varsigma) \strongsim_{\trg{p}}
 (\trg{s_2}, \varsigma)
\wedge
(\trg{s_1}, \varsigma) 
~\trg{\xrightharpoonup{\neutralcol{\tau}}_{
        \trg{p}}}~
(\trg{s_1'}, \varsigma)
~\implies~
\exists \trg{s_2'}.~(\trg{s_2}, \varsigma)
~\trg{\xrightharpoonup{\neutralcol{\tau}}_{
        \trg{p}}}~
(\trg{s_2'}, \varsigma)
\wedge
(\trg{s_1'}, \varsigma) \strongsim_{\trg{p}}
(\trg{s_2'}, \varsigma)
$
\end{lemma}

\begin{lemma}[Strong similarity is weakened by an output
    action]\label{assm-weakening}
    $ 
    \lambda \in \overset{\bullet}{!} \wedge
    (\trg{s_1}, \varsigma) \strongsim_{\trg{p}}
    (\trg{s_2}, \varsigma)
    \wedge
    (\trg{s_1}, \varsigma) 
    ~\trg{\xrightharpoonup{\neutralcol{\lambda}}_{
            \trg{p}}}~
    (\trg{s_1'}, \varsigma')
    ~\implies~
    \exists \trg{s_2'}.~(\trg{s_2}, \varsigma)
    ~\trg{\xrightharpoonup{\neutralcol{\lambda}}_{
            \trg{p}}}~
    (\trg{s_2'}, \varsigma')
    \wedge
    (\trg{s_1'}, \varsigma') \weaksim_{\trg{p}}
    (\trg{s_2'}, \varsigma')
    $
\end{lemma}

\begin{lemma}[Option simulation of weak similarity]
    \label{assm-option-simulation}
$ 
(\trg{s_1}, \varsigma) \weaksim_{\trg{p}}
(\trg{s_2}, \varsigma)
\wedge
(\trg{s_1}, \varsigma) 
~\trg{\xrightharpoonup{\neutralcol{\tau}}_{
        \trg{p}}}~
(\trg{s_1'}, \varsigma)
~\implies~
(\trg{s_1'}, \varsigma) \weaksim_{\trg{p}}
(\trg{s_2}, \varsigma)
$
\end{lemma}

\begin{lemma}[Weak similarity is strengthened by aligned input
    actions]\label{assm-strengthening}
   $ 
    \lambda \in \overset{\bullet}{?} \wedge
    (\trg{s_1}, \varsigma) \weaksim_{\trg{p}}
    (\trg{s_2}, \varsigma)
    \wedge
    (\trg{s_1}, \varsigma) 
    ~\trg{\xrightharpoonup{\neutralcol{\lambda}}_{
            \trg{p}}}~
    (\trg{s_1'}, \varsigma')
    \wedge
    (\trg{s_2}, \varsigma) 
    ~\trg{\xrightharpoonup{\neutralcol{\lambda}}_{
            \trg{p}}}~
    (\trg{s_2'}, \varsigma')
    ~\implies~
    (\trg{s_1'}, \varsigma') \strongsim_{\trg{p}}
    (\trg{s_2'}, \varsigma')
    $
\end{lemma}

%

\begin{lemma}[Adequacy of $\ms{emulate\_invariants}$]
    \label{assm-emul-adequacy}
~\\$ 
\fun{emulate\_invariants}{\src{s}}_{\alpha, i, \src{p}} ~\wedge~
\alpha(i) \in \overset{\bullet}{?} \cup \{\checkmark\}
~\implies~ \exists \src{s'}.~(\src{s}, \_)
~\src{\paronto{\neutralcol{\alpha(i)}}_{p}}~
(\src{s'}, \_)$
\end{lemma}

\begin{lemma}[Preservation of $\ms{emulate\_invariants}$]
    \label{assm-emul-invar}
~\\$ 
\fun{emulate\_invariants}{\src{s}}_{\alpha, i, \src{p}} \wedge
(\src{s}, \_)
~\src{\paronto{\neutralcol{\alpha(i)}}_{p}}~
(\src{s'}, \_)
~\implies~
\fun{emulate\_invariants}{\src{s'}}_{\alpha, i + 1, \src{p}}
$
\end{lemma}

Using these lemmas, we prove \Cref{tricl-simulation-lemma} as follows.

\begin{proof}(of \Cref{tricl-simulation-lemma}; simplified)
	~\\\underline{Hypotheses (unfolding \Cref{tricl-def}):}
	\begin{enumerate}
		\item \label{assm-alpha-alt} $\alpha \in \mathtt{Alt}$
		\item \label{assm-compiler-rel}
		$\src{s_{emu}} \statecrossrelated_{\src{p_s}} \trg{s_{med}}$
		\item \label{assm-emulate-inv}
		$\fun{emulate\_invariants}{\src{s_{emu}}}_{\alpha, i,
			\src{p_s}}$
		\item \label{assm-!-implies}
		$\alpha(i) \in \overset{\bullet}{!} 
		\implies
		(\trg{s_{med}}, \varsigma)
		\strongsim_{\compilersymbol{\src{p_s}}} 
		(\trg{s_{given}}, \varsigma)$
		\item \label{assm-?-implies}
		$\alpha(i) \in \overset{\bullet}{?} \implies
		(\trg{s_{med}}, \varsigma) \weaksim_{\compilersymbol{\src{p_s}}} 
		(\trg{s_{given}}, \varsigma)$
		
		\item \label{assm-given-step} $(\trg{s_{given}}, \varsigma) 
		~\trg{\paronto{\neutralcol{\alpha(i)}}_{\compilersymbol{\src{p_s}}}}~
		(\trg{s_{given}'}, \varsigma')$
	\end{enumerate}
	
	We consider two cases for the trace step $\alpha(i)$, which we obtain by unfolding
	the definition of $\mathtt{Alt}$ in hypothesis \ref{assm-alpha-alt}
	(we ignore in this proof sketch
	the case of $\alpha(i) = \checkmark$):
	\begin{itemize}
		\item \underline{Case $\alpha(i) \in \overset{\bullet}{?}$: (\textbf{4 subgoals})}
		\begin{enumerate}[label=(\alph*)]
			\item \label{subgoal-src-step} Obtain $\src{s_{emu}'}$ where $(\src{s_{emu}}, \varsigma)
					~\src{\paronto{\neutralcol{\alpha(i)}}_{\compilersymbol{\src{p_s}}}}~
					(\src{s_{emu}'}, \varsigma')$ as follows:
			Instantiate \Cref{assm-emul-adequacy} (adequacy of the emulation invariants)
			with both hypothesis \ref{assm-emulate-inv}
			and the case condition to obtain \src{s_{emu}'} that satisfies
			$(\src{s_{emu}}, \varsigma)
			~\src{\paronto{\neutralcol{\alpha(i)}}_{\compilersymbol{\src{p_s}}}}~
			(\src{s_{emu}'}, \varsigma'')$ for some $\varsigma''$.
			Then, by inversion of this step (and unfolding the computation of the informative label),
			we know $\varsigma'' = \varsigma'$.
			
			\item \label{subgoal-med-steps} Obtain $\trg{s_{med}'}$ where $(\trg{s_{med}}, \varsigma) 
					~\trg{\paronto{\neutralcol{\alpha(i)}}_{\compilersymbol{\src{p_s}}}}~
					(\trg{s_{med}'}, \varsigma')$ and
					$\src{s_{emu}'} \statecrossrelated_{\src{p_s}} \trg{s_{med}'}$
					as follows:
			Instantiate \Cref{assm-fwdsim} with both hypothesis
			\ref{assm-compiler-rel}
			and statement \ref{subgoal-src-step} from above.
			
			\item \label{subgoal-med-strong-given}
                        Obtain strong similarity $(\trg{s_{med}'}, \varsigma)
					\strongsim_{\compilersymbol{\src{p_s}}} 
					(\trg{s_{given}'}, \varsigma)$ as follows:
			First, unfold both statement \ref{subgoal-med-steps} and hypothesis \ref{assm-given-step}
			to obtain respectively
			the two states \trg{s_{med}''} and \trg{s_{given}''} just before the border crossing.
			Now instantiate \Cref{assm-option-simulation} (option simulation) with 
			hypothesis \ref{assm-?-implies}
			twice, once with \trg{s_{med}''} for \trg{s_1'} and once with \trg{s_{given}''}
			for \trg{s_1'} (use symmetry of weak similarity before the second instantiation).
			Now (by transitivity of weak similarity), we know that $(\trg{s_{med}''}, \varsigma)
			\weaksim_{\compilersymbol{\src{p_s}}} (\trg{s_{given}''}, \varsigma)$.
			Thus, instantiate \Cref{assm-strengthening} (strengthening) to obtain
			$(\trg{s_{med}'}, \varsigma)
			\strongsim_{\compilersymbol{\src{p_s}}} 
			(\trg{s_{given}'}, \varsigma)$.
			
			\item Obtain $\fun{emulate\_invariants}{\src{s_{emu}'}}_{\alpha, i+1,
						\src{p_s}}$ as follows:
			Instantiate \Cref{assm-emul-invar} (preservation of the emulation invariants)
			with hypothesis \ref{assm-emulate-inv} and statement \ref{subgoal-src-step} from
			above.
		\end{enumerate}
		
                
		\item \underline{Case $\alpha(i) \in \overset{\bullet}{!}$: (\textbf{3 subgoals})}
		
		\begin{enumerate}[label=(\alph*)]
			\item \label{subgoal-med-steps-2} 
			Obtain $\trg{s_{med}'}$ where $(\trg{s_{med}}, \varsigma) 
					~\trg{\paronto{\neutralcol{\alpha(i)}}_{\compilersymbol{\src{p_s}}}}~
					(\trg{s_{med}'}, \varsigma')$ and
					$(\trg{s_{med}'}, \varsigma) \weaksim_{\compilersymbol{\src{p_s}}} 
					(\trg{s_{given}'}, \varsigma)$
					as follows:
			First, unfold the definition of the given step of hypothesis \ref{assm-given-step}
			to obtain star-many $\tau$ steps leading to some state \trg{s_{given}''} where
			also $(\trg{s_{given}''}, \varsigma)~\trg{\xrightharpoonup{\neutralcol{\alpha(i)}}_{
					\neutralcol{\compilersymbol{\src{p_s}}}}}~(\trg{s_{given}'}, \varsigma')$.
			Now instantiate (the star version of) \Cref{assm-lock-step-simulation} 
			(lock-step simulation) with both hypothesis \ref{assm-!-implies} (after applying
			symmetry of strong similarity)
			and state \trg{s_{given}''} (the $\tau$ steps) to obtain
			some \trg{s_{med}''} where $(\trg{s_{given}''}, \varsigma)
			\strongsim_{\compilersymbol{\src{p_s}}}
			(\trg{s_{med}''}, \varsigma)$.
			Next, use this 
			together with the step $(\trg{s_{given}''}, \varsigma)~\trg{\xrightharpoonup{\neutralcol{\alpha(i)}}_{
					\neutralcol{\compilersymbol{\src{p_s}}}}}~(\trg{s_{given}'}, \varsigma')$,
			which we obtained above,
			to instantiate
			\Cref{assm-weakening} (weakening).
			Next,
			apply symmetry of weak similarity to obtain the subgoal.
			
			\item Obtain \src{s_{emu}'} where $(\src{s_{emu}}, \varsigma)
					~\src{\paronto{\neutralcol{\alpha(i)}}_{\compilersymbol{\src{p_s}}}}~
					(\src{s_{emu}'}, \varsigma')$ and
					$\src{s_{emu}'} \statecrossrelated_{\src{p_s}} \trg{s_{med}'}$ as follows:
			Instantiate \Cref{assm-bckwdsim} (backward simulation) using both hypothesis
			\ref{assm-compiler-rel} and the step of \trg{s_{med}} that we obtained in
			the previous subgoal.
			
			\item Obtain $\fun{emulate\_invariants}{\src{s_{emu}'}}_{\alpha, i+1,
						\src{p_s}}$ as follows:
			Instantiate \Cref{assm-emul-invar} (preservation of the emulation invariants)
			with both hypothesis \ref{assm-emulate-inv} and the step of \src{s_{emu}}
			that we obtained in the previous subgoal.
		\end{enumerate}
%
%
	\end{itemize}
\end{proof}



\end{document}